\documentclass[12pt]{cernprep}
\usepackage{graphicx}
\usepackage{amssymb}
\usepackage{epsfig,color,rotating}
\begin{document}
\newcommand{\dedx}{\mbox{${\rm d}E/{\rm d}x$}}
\newcommand{\EcB}{$E \! \times \! B$}
\newcommand{\omt}{$\omega \tau$}
\newcommand{\omtsq}{$(\omega \tau )^2$}
\newcommand{\rphi}{\mbox{$r \! \cdot \! \phi$}}
\newcommand{\srphi}{\mbox{$\sigma_{r \! \cdot \! \phi}$}}
\newcommand{\dg}{\mbox{`durchgriff'}}
\newcommand{\mg}{\mbox{`margaritka'}}
\newcommand{\pT}{\mbox{$p_{\rm T}$}}
\newcommand{\GeVc}{\mbox{GeV/{\it c}}}
\newcommand{\MeVc}{\mbox{MeV/{\it c}}}
\def\kr{$^{83{\rm m}}$Kr\ }
\begin{titlepage}
\docnum{CERN--PH--EP/2010--017}
\date{7 May 2010} 
\vspace{1cm}
\title{\Large HARP--CDP hadroproduction data: \\
Comparison with FLUKA and GEANT4 simulations}

\begin{abstract}
We report on the comparison of production characteristics of secondary protons and charged
pions in the interactions of protons and charged pions with momentum between 3~GeV/{\it c} and
15~GeV/{\it c} with beryllium, copper, and tantalum nuclei, with simulations by the FLUKA and Geant4 Monte Carlo tool kits.
Overall production cross-sections are reasonably well reproduced, within factors of two.
In more detail, there are areas with poor agreement that are unsatisfactory and
call for modelling improvements. Overall, the
current FLUKA simulation fares better than the current Geant4 simulation.
\end{abstract}

\vfill  \normalsize
\begin{center}
The HARP--CDP group  \\  

\vspace*{2mm} 

A.~Bolshakova$^1$, 
I.~Boyko$^1$, 
G.~Chelkov$^{1a}$, 
D.~Dedovitch$^1$, 
A.~Elagin$^{1b}$, 
D.~Emelyanov$^1$,
M.~Gostkin$^1$,
A.~Guskov$^1$, 
Z.~Kroumchtein$^1$, 
Yu.~Nefedov$^1$, 
K.~Nikolaev$^1$, 
A.~Zhemchugov$^1$, 
F.~Dydak$^2$, 
J.~Wotschack$^{2*}$, 
A.~De~Min$^{3c}$,
V.~Ammosov$^{4\dagger}$, 
V.~Gapienko$^4$, 
V.~Koreshev$^4$, 
A.~Semak$^4$, 
Yu.~Sviridov$^4$, 
E.~Usenko$^{4d}$, 
V.~Zaets$^4$ 
\\
 
\vspace*{5mm} 

$^1$~{\bf Joint Institute for Nuclear Research, Dubna, Russia} \\
$^2$~{\bf CERN, Geneva, Switzerland} \\ 
$^3$~{\bf Politecnico di Milano and INFN, 
Sezione di Milano-Bicocca, Milan, Italy} \\
$^4$~{\bf Institute of High Energy Physics, Protvino, Russia} \\

\vspace*{5mm}

\submitted{(To be submitted to Eur. Phys. J. C)}
\end{center}

\vspace*{5mm}
\rule{0.9\textwidth}{0.2mm}

\begin{footnotesize}

$^a$~Also at the Moscow Institute of Physics and Technology, Moscow, Russia 

$^b$~Now at Texas A\&M University, College Station, USA 

$^c$~On leave of absence at 
Ecole Polytechnique F\'{e}d\'{e}rale, Lausanne, Switzerland 

$^d$~Now at Institute for Nuclear Research RAS, Moscow, Russia

$^{\dagger}$~Deceased on 11 January 2010

$^*$~Corresponding author; e-mail: joerg.wotschack@cern.ch
\end{footnotesize}

\end{titlepage}


\newpage

\section{Introduction}

The HARP experiment arose from the realization that the 
inclusive differential cross-sections of hadron production 
in the interactions of few GeV/{\it c} protons with nuclei were 
known only within a factor of two to three, while 
more precise cross-sections are in demand for several 
reasons. Among them are the understanding of the underlying 
physics and the modelling of Monte Carlo generators of 
hadron--nucleus collisions.

The HARP experiment 
was designed to carry 
out a programme of systematic and precise 
(i.e., at the few per cent level) measurements of 
hadron production by protons and pions with momenta from 
1.5 to 15~GeV/{\it c}, on a variety of target nuclei ranging from hydrogen to lead.

The HARP detector combined a forward spectrometer with a 
large-angle spectrometer. The latter comprised a 
cylindrical Time Projection 
Chamber (TPC) around the target and an array of 
Resistive Plate Chambers (RPCs) that surrounded the 
TPC. The purpose of the TPC was track 
reconstruction and particle identification by \dedx . The 
purpose of the RPCs was to complement the 
particle identification by time of flight.

The HARP experiment took data at the CERN Proton Synchrotron 
in 2001 and 2002.

Several papers reported on the measurement of 
inclusive cross-sections of large-angle production (polar angle $\theta$ in the 
range $20^\circ < \theta < 125^\circ$) of secondary protons and charged pions,
in the interactions with 5\% $\lambda_{\rm int}$ beryllium, copper, tantalum and lead targets
of protons and pions with beam momenta of $\pm3.0$, 
$\pm5.0$, $\pm8.0$ ($+8.9$ for beryllium), $\pm12.0$, and $\pm15.0$~GeV/{\it c}~\cite{Beryllium1,Beryllium2,Copper,
Tantalum,Lead}. Besides, one paper~\cite{GEANTpub} reported on disagreements 
between data and simulations by the Geant4 Monte Carlo tool kit~\cite{Geant4}, which led to
significant improvements in its simulation code~\cite{ResponsetoGEANTpub}.

In this paper, we report in more detail on comparisons of hadroproduction data with 
simulations by the FLUKA~\cite{FLUKA} and Geant4 Monte Carlo tool kits. 

Our work involves only the HARP large-angle spectrometer. The data analysis 
is based on our calibrations of the HARP TPC and RPCs published in 
Refs.~\cite{TPCpub,RPCpub}.

\section{The beams and the HARP spectrometer}

The protons and pions were delivered by
the T9 beam line in the East Hall of CERN's Proton Synchrotron.
This beam line supports beam momenta between 1.5 and 15~GeV/{\it c},
with a momentum bite $\Delta p/p \sim 1$\%.

The beam instrumentation, the definition of the beam particle trajectory,
the cuts to select `good' beam particles, and the muon and electron contaminations
of the particle beams, are the same as described, e.g., 
in Ref.~\cite{Copper}.
The targets were discs made of 
high-purity material,  5\% $\lambda_{\rm int}$ thick. 

The momentum resolution $\sigma (1/p_{\rm T})$ of the HARP--TPC 
is typically 0.2~(GeV/{\it c})$^{-1}$ 
and worsens towards small relative particle
velocity $\beta$ and small polar angle $\theta$.
The absolute momentum scale is determined to be correct to 
better than 2\%, both for positively and negatively
charged particles.
 
The polar angle $\theta$ is measured in the TPC with a 
resolution of $\sim$13~mrad, for a proton with 
$p_{\rm T} = 500$~MeV/{\it c} in the TPC gas and a polar  
angle of $\theta = 60^\circ$. 
The polar-angle scale is correct to better than 2~mrad.     

The TPC measures \dedx\ with a resolution of 16\% for a 
track length of 300~mm.

The intrinsic efficiency of the RPCs that surround 
the TPC is better than 98\%.

The intrinsic time resolution of the RPCs is 127~ps and
the system time-of-flight resolution (that includes the
jitter of the arrival time of the beam particle at the target)
is 175~ps. 

To separate measured particles into species, we
assign on the basis of \dedx\ and $\beta$ to each particle a 
probability of being a proton,
a pion (muon), or an electron, respectively. The probabilities
add up to unity, so that the number of particles is conserved.
These probabilities are used for weighting when entering 
tracks into plots or tables.

\section{Modus operandi of the comparison between data and simulations}

All data shown in this paper have been published~\cite{Beryllium1,Beryllium2,Copper,
Tantalum,Lead} in the form of double-differential
inclusive cross-sections ${\rm d}^2 \sigma / {\rm d}p {\rm d}\Omega$
[mbarn/(sr$\cdot$GeV/{\it c})],
in the transverse-momentum range $0.10 < p_{\rm T} < 1.25$ GeV/{\it c} and the
polar-angle range $20^\circ < \theta < 125^\circ$. For ease of use, the cross-sections are also available in
computer-readable form as ASCII files~\cite{Beryllium1tables,Beryllium2tables,Coppertables,
Tantalumtables,Leadtables}. 

For the comparison with 
simulations, cross-sections are integrated over two regions: the `intermediate-angle' region
($20^\circ < \theta < 50^\circ$) and the `large-angle' region ($50^\circ < \theta < 125^\circ$).
For $\pi^+$ and $\pi^-$ secondaries, the $p_{\rm T}$ ranges are 
$0.10 < p_{\rm T} < 0.72$ GeV/{\it c} in the intermediate-angle region, and 
$0.16 < p_{\rm T} < 1.25$ GeV/{\it c} in the large-angle region. For proton 
secondaries, the $p_{\rm T}$ range is
$0.30 < p_{\rm T} < 0.72$ GeV/{\it c}, for the large energy loss by ionization in the target and in materials 
before entering the active TPC volume. Cross-sections of protons are given
only in the intermediate-angle region because the minimum $p_{\rm T}$ of
protons in the large-angle region is even larger, about twice the one in the
intermediate-angle region.

The measured differential cross-sections in the said regions were obtained 
by integrating over respective bins of polar angle and transverse momentum, 
taking into account the correlation of systematic errors.
The uncertainties of the shown inclusive cross-sections
are at the level of 3\%. The contribution of statistical
errors is negligible, with the exception of the cross-sections for 15~GeV/c $\pi^+$
beams. The largest contributions to the systematic error arise from the overall 
normalization (2\%) and from the uncertainty
of the momentum scale of secondaries (2\%).

The Monte Carlo tool kits FLUKA and Geant4\footnote{We used the program versions
FLUKA 2008.3c and Geant 4.9.3.} are run with protons and charged-pion beams
with the same beam momenta that interact with beryllium, copper
and tantalum target nuclei. Only final-state hadrons that stem from the target nuclei are
taken into account. For the same regions selected for the presentation
of data, integrated inclusive cross-sections are extracted.

The comparison is made separately (i) for the intermediate-angle and the large-angle
regions, (ii) for secondary protons, $\pi^+$'s and $\pi^-$'s, and (iii) for incoming beam
protons, $\pi^+$'s and $\pi^-$'s in the momentum range 3--15~GeV/{\it c}. 

\begin{figure}[htp]
\begin{center}
\begin{tabular}{cc}
\includegraphics[width=0.50\textwidth]{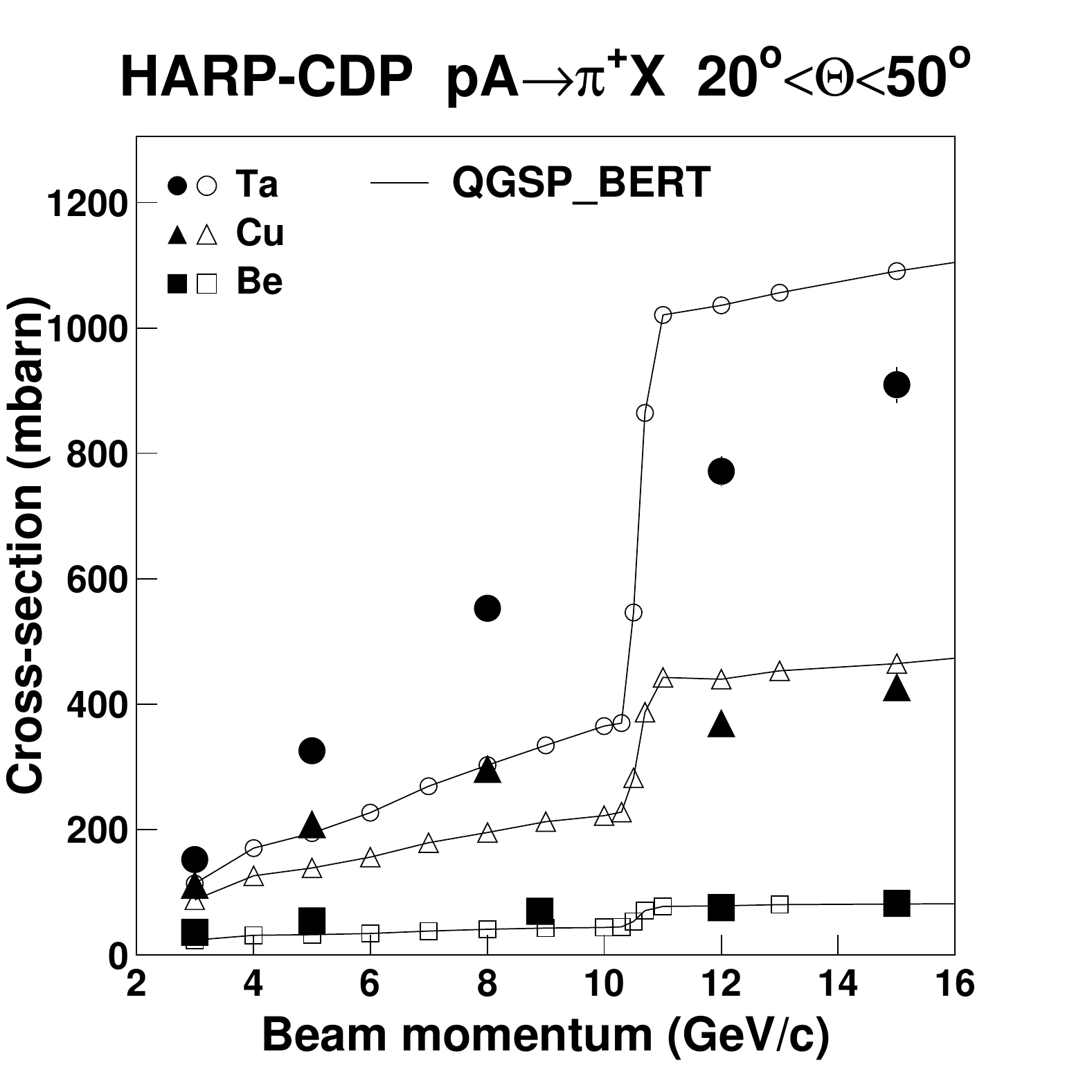} &
\includegraphics[width=0.50\textwidth]{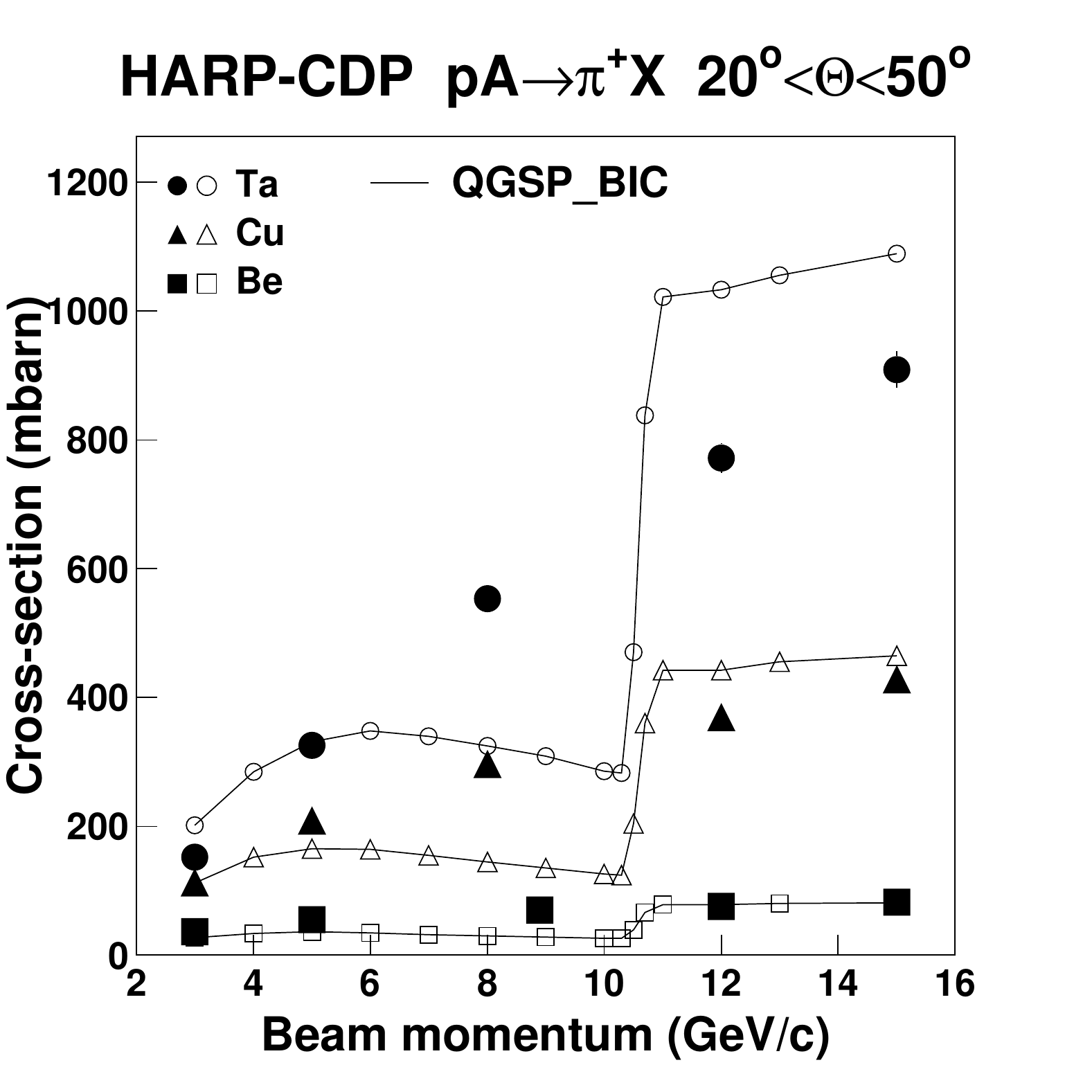} \\
\includegraphics[width=0.50\textwidth]{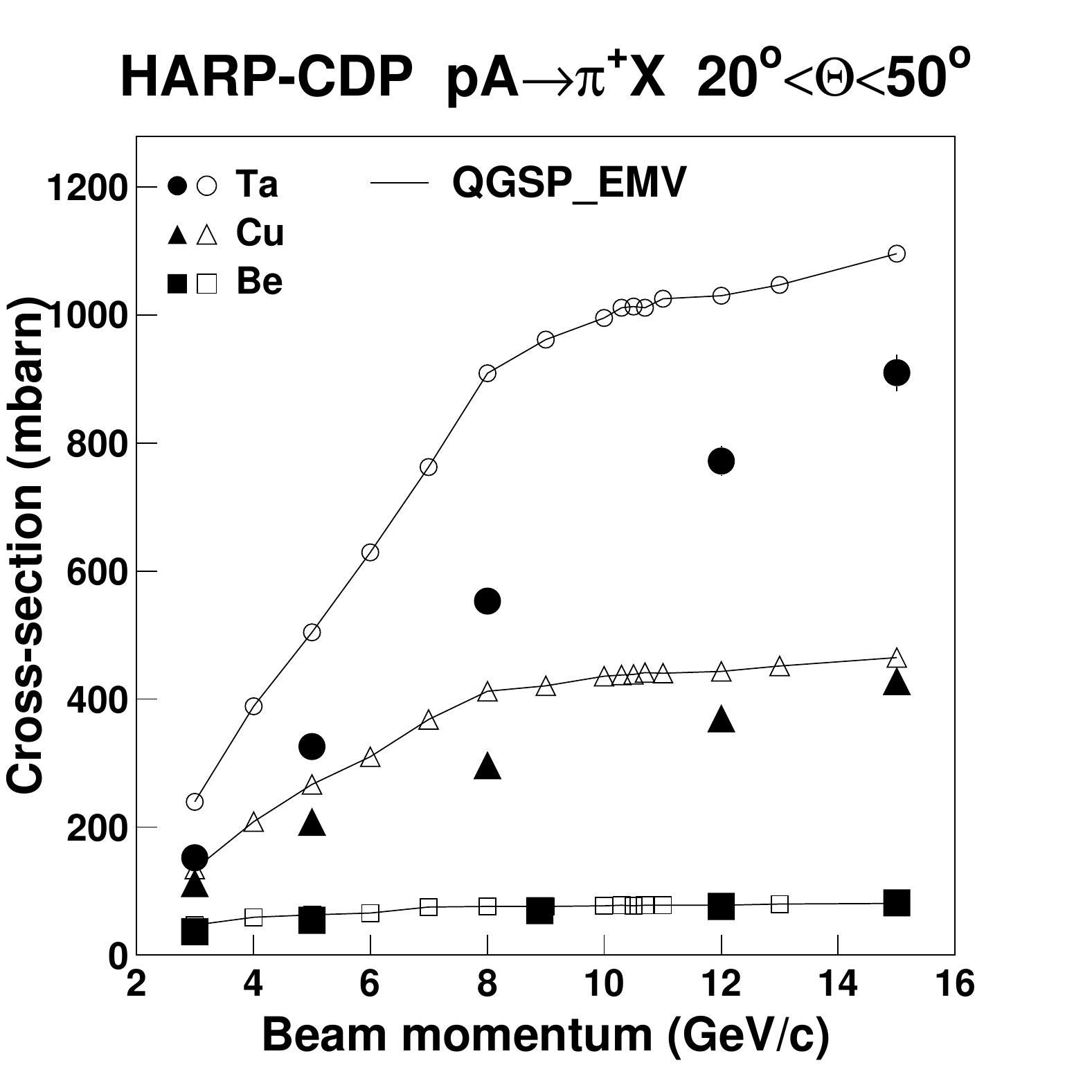} &
\includegraphics[width=0.50\textwidth]{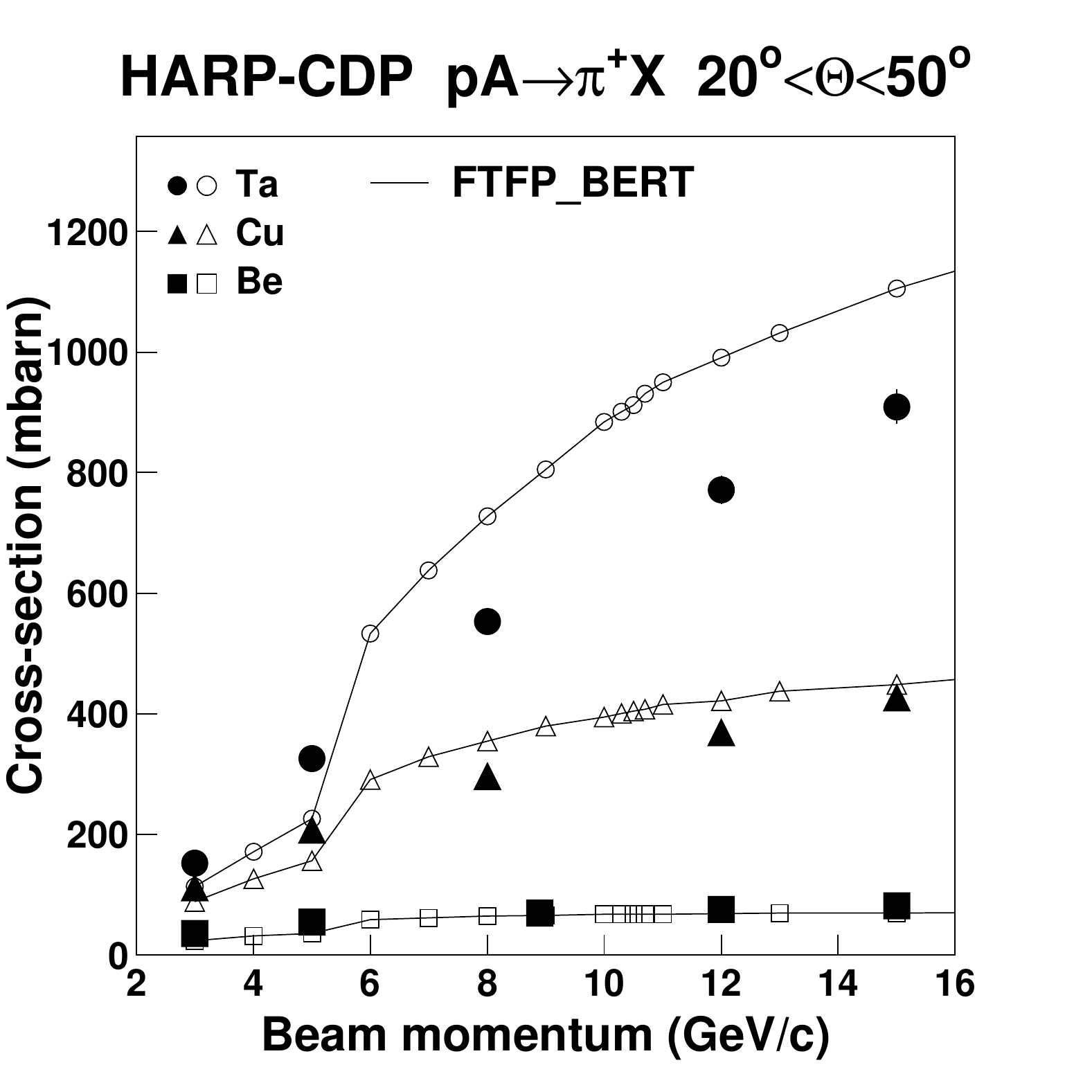} \\
\end{tabular}
\caption{Comparison of measured inclusive $\pi^+$ production cross-sections by protons on Be (squares),
Cu (triangles) and Ta (circles) with Geant4 simulations employing the  QGSP\_BERT 
(upper left panel), QGSP\_BIC (upper right panel), QGSP\_EMV (lower left panel) and 
FTFP\_BERT (lower right panel), in the
intermediate-angle region, as a
function of beam momentum; the data
are shown with black symbols, the simulations with open symbols; here and in further
similar figures, the simulated points are connected by lines to guide the eye.}
\label{comparisonofG4physicslists}
\end{center}
\end{figure}

For Geant4, the QGSP\_BERT `physics list' was selected for being the preferred choice
of the LHC Collaborations ATLAS and CMS~\cite{PreferredChoice}. In order to assess
the differences with other popular physics lists,  in 
Fig.~\ref{comparisonofG4physicslists}
QGSP\_BERT simulations of inclusive $\pi^+$ production by protons on beryllium, copper and tantalum nuclei are compared with the simulations 
employing the Geant4 physics lists QGSP\_BIC, QGSP\_EMV and FTFP\_BERT.
There are remarkable differences between the simulations, similar in 
size as the differences between the data and, e.g., the
QGSP\_BERT simulation. 
Most of these differences show up for beam proton momenta below 10 GeV/{\it c}. 
Between 10 and 15 GeV/{\it c} beam proton momentum, the chosen four physics lists 
give comparable results.

Our aim is not to test all possible Geant4 physics lists against our data. Rather, we
wish to point to areas, using the QGSP\_BERT physics list as an example, where 
data and simulation seriously disagree.Ê

We trust that our data are useful to the simulation developers and provide 
guidance for modelling improvements.

The comparison of data with FLUKA and Geant4 QGSP\_BERT simulations
is concentrated in Section~\ref{Comparisonwithdata}. A critical appraisal
is found in Section~\ref{Appraisal}.

\section{Comparison of data with FLUKA and Geant4 simulations}
\label{Comparisonwithdata}

\begin{figure}[htp]
\begin{center}
\begin{tabular}{c}
\includegraphics[width=0.45\textwidth]{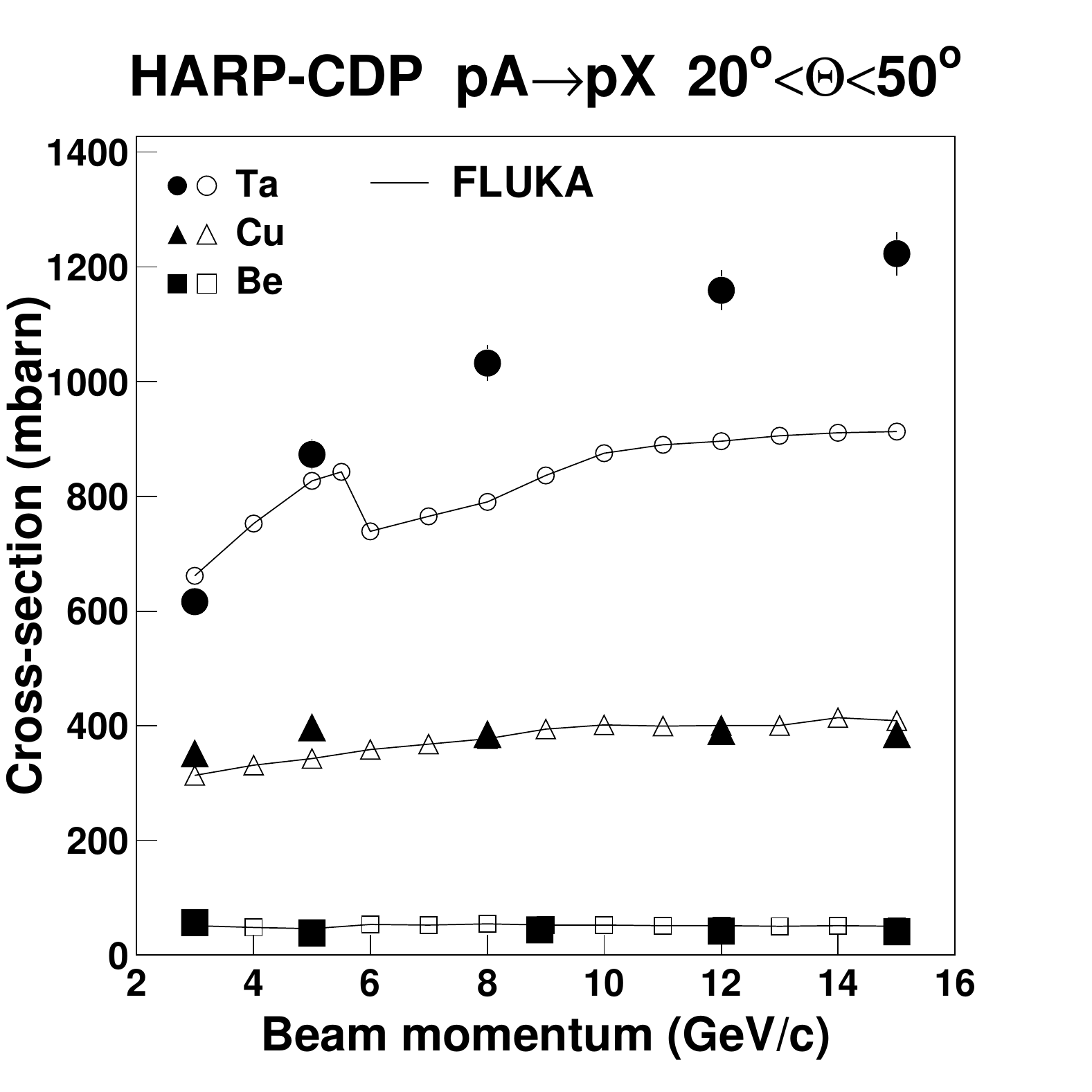} \\
\includegraphics[width=0.45\textwidth]{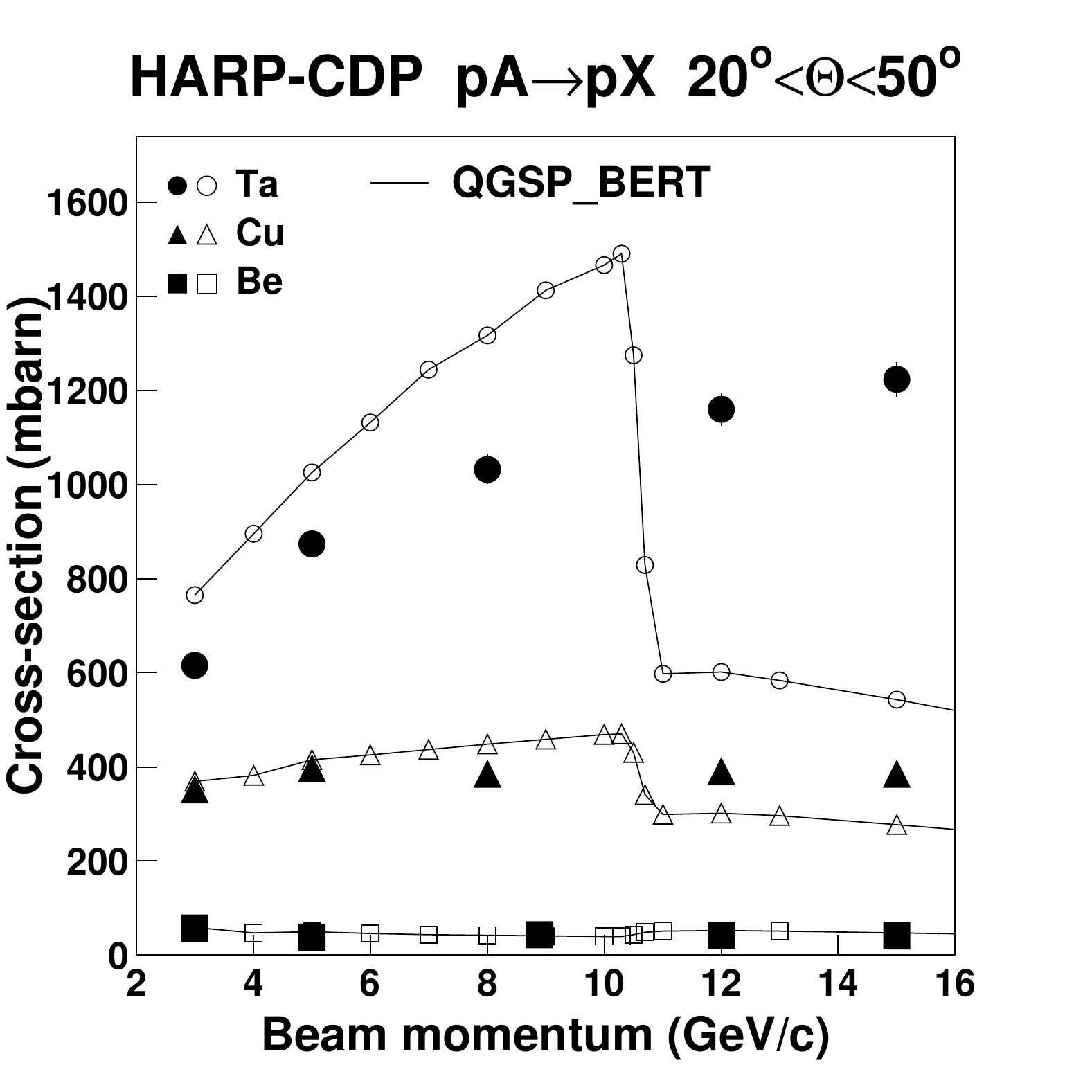} \\
\end{tabular}
\caption{Comparison of measured (black symbols) inclusive proton production
cross-sections by protons on Be, Cu and Ta nuclei,  in the intermediate-angle region, as a
function of beam momentum, with FLUKA (upper panel) and Geant4 (lower panel) 
simulations (open symbols).} 
\label{FLUKAG4protonsbyprotons-int}
\end{center}
\end{figure}

\begin{figure}[h]
\begin{center}
\begin{tabular}{cc}
\includegraphics[width=0.45\textwidth]{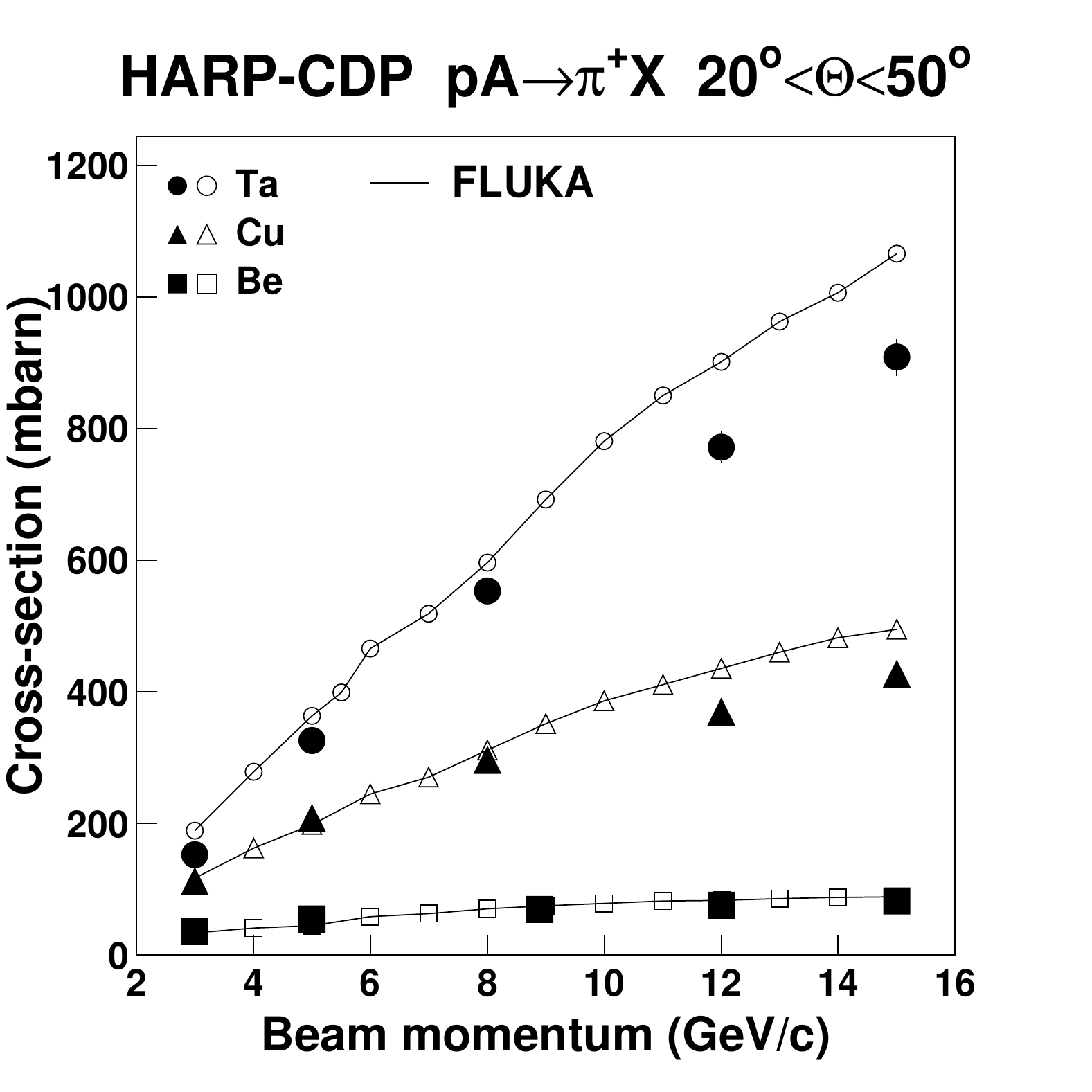} &
\includegraphics[width=0.45\textwidth]{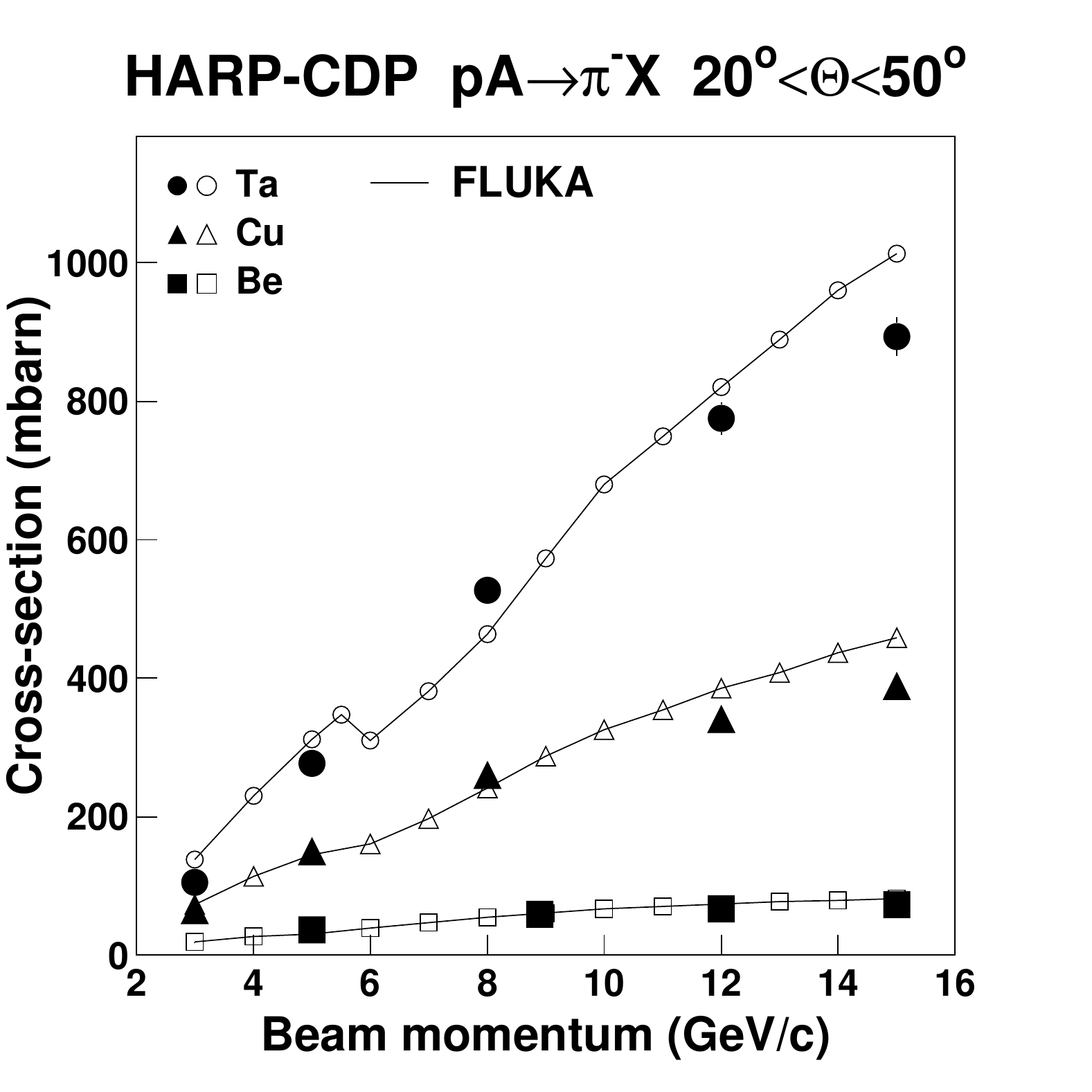} \\
\includegraphics[width=0.45\textwidth]{bert_pro_pip_fwd.pdf} &
\includegraphics[width=0.45\textwidth]{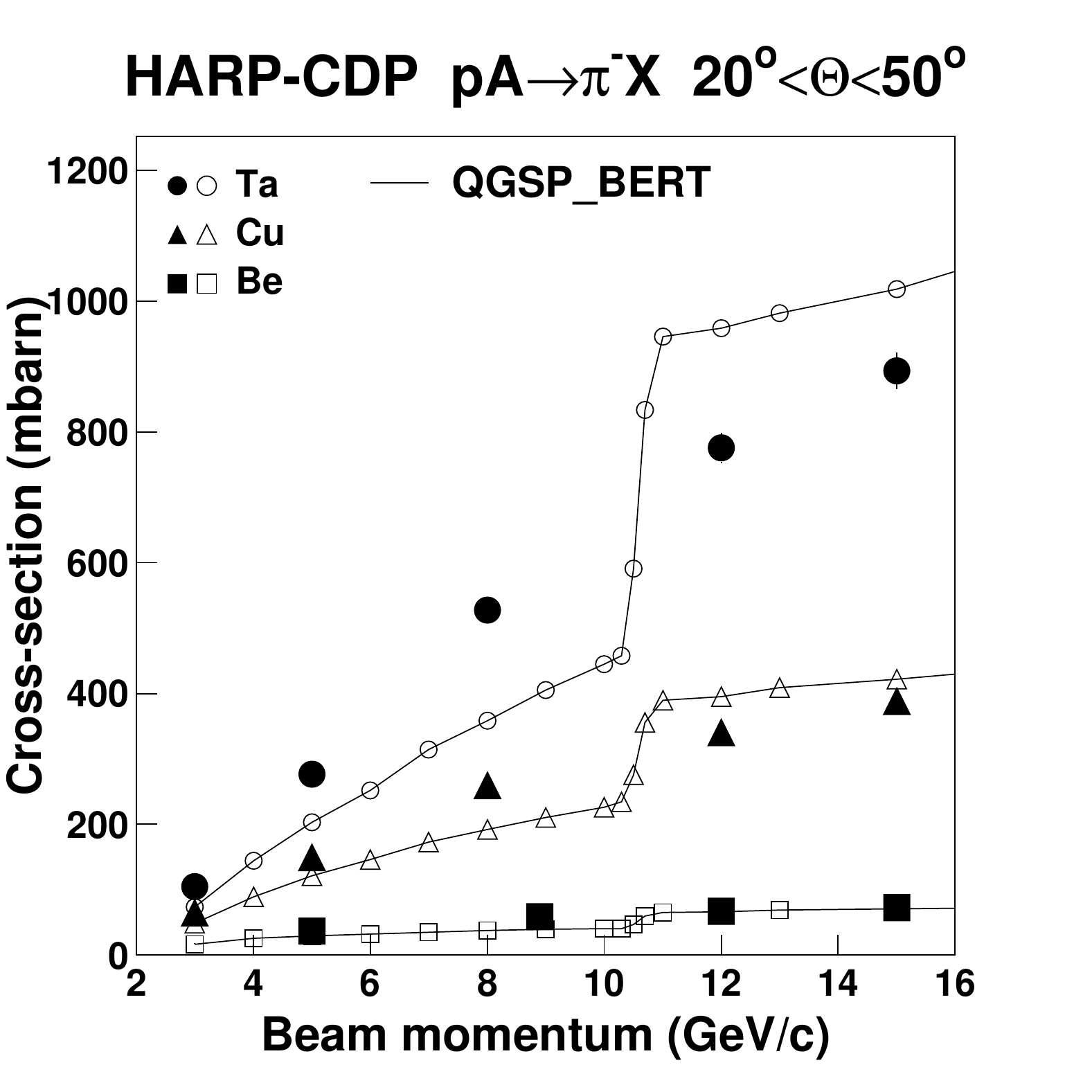} \\
\end{tabular}
\caption{Comparison of measured inclusive $\pi^+$ (left panels) and 
$\pi^-$ (right panels) production cross-sections
by protons on Be, Cu and Ta nuclei (black symbols),  in the intermediate-angle region, 
with FLUKA and Geant4 simulations (open symbols).} 
\label{FLUKAG4pionsbyprotons-int}
\end{center}
\end{figure}

\begin{figure}[h]
\begin{center}
\begin{tabular}{cc}
\includegraphics[width=0.45\textwidth]{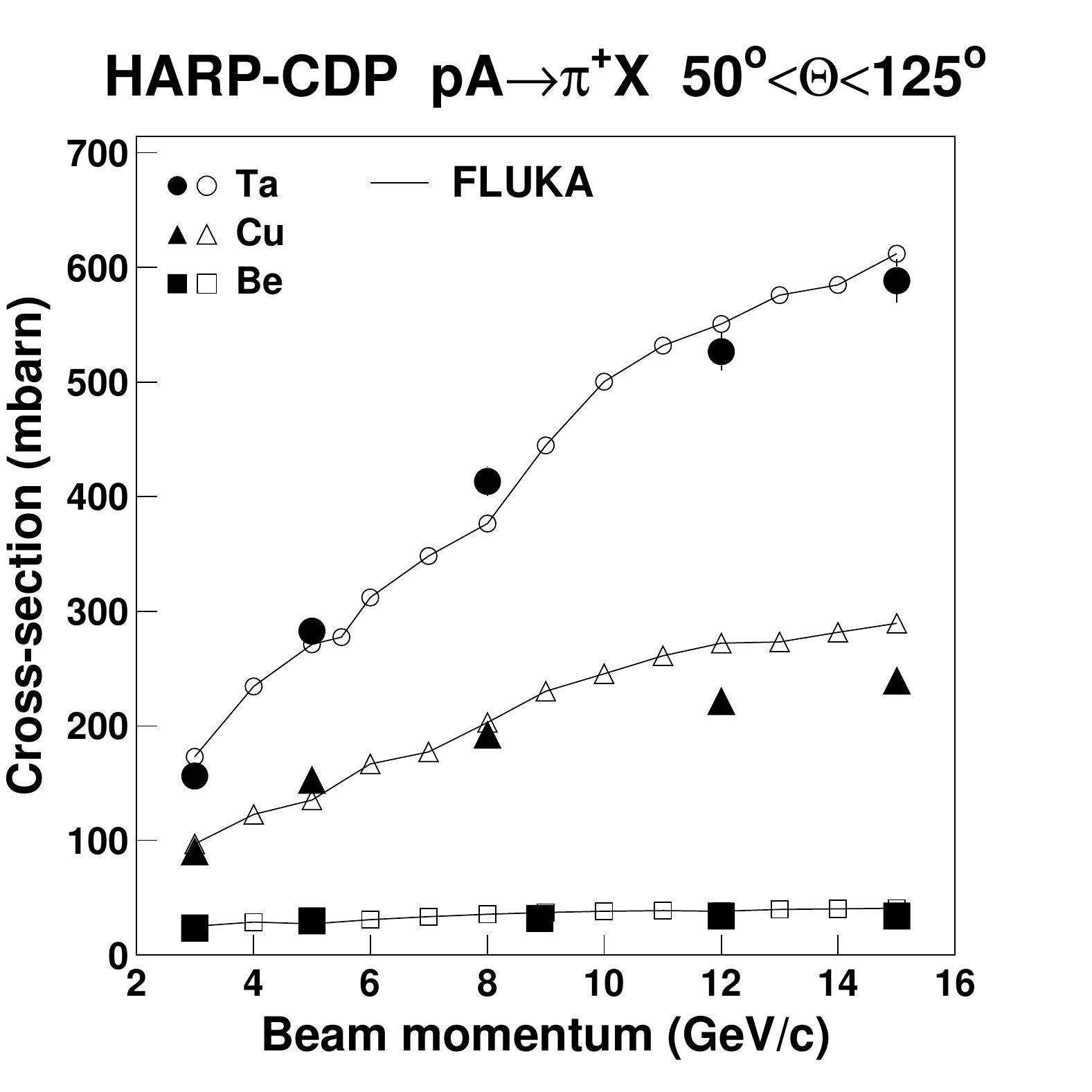} &
\includegraphics[width=0.45\textwidth]{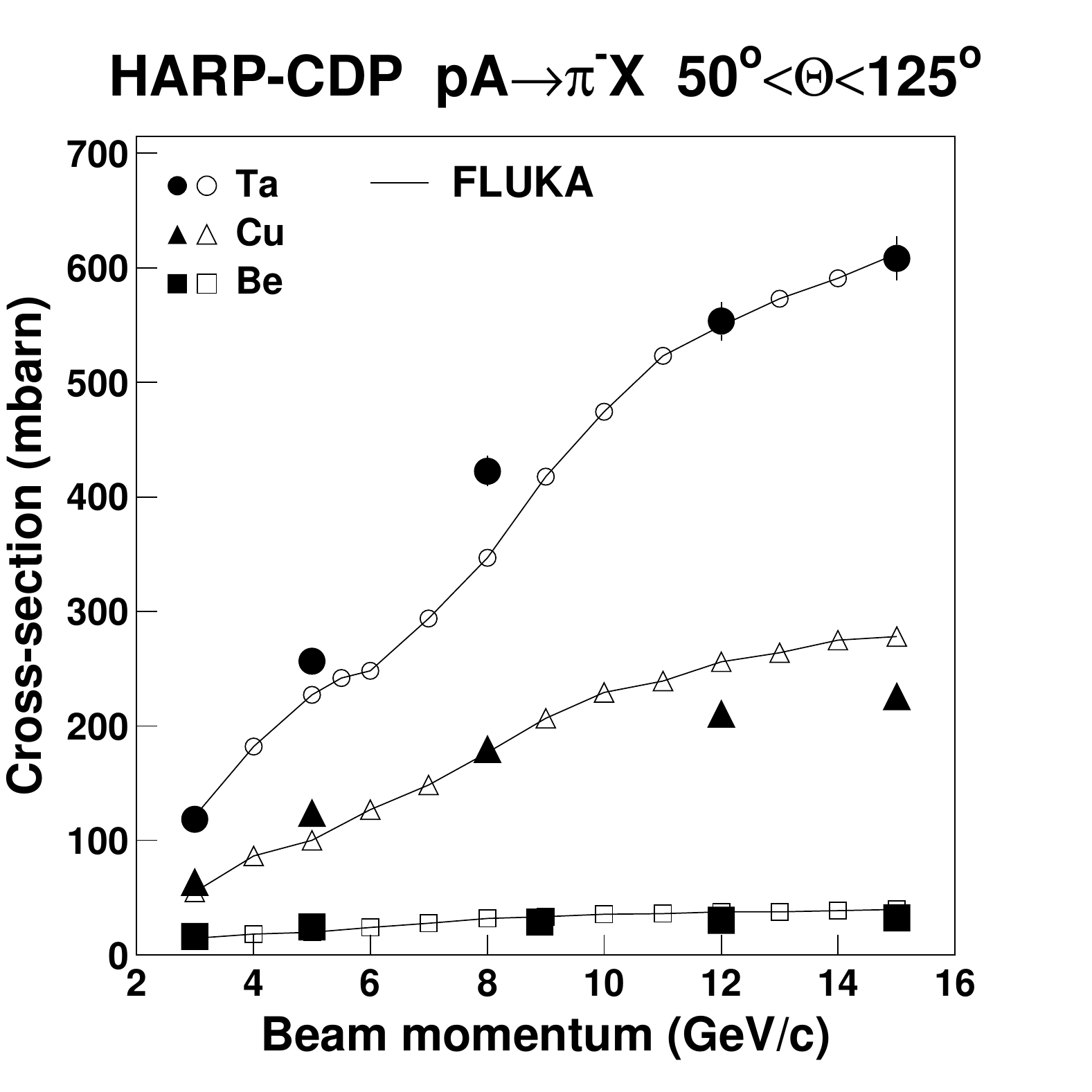} \\
\includegraphics[width=0.45\textwidth]{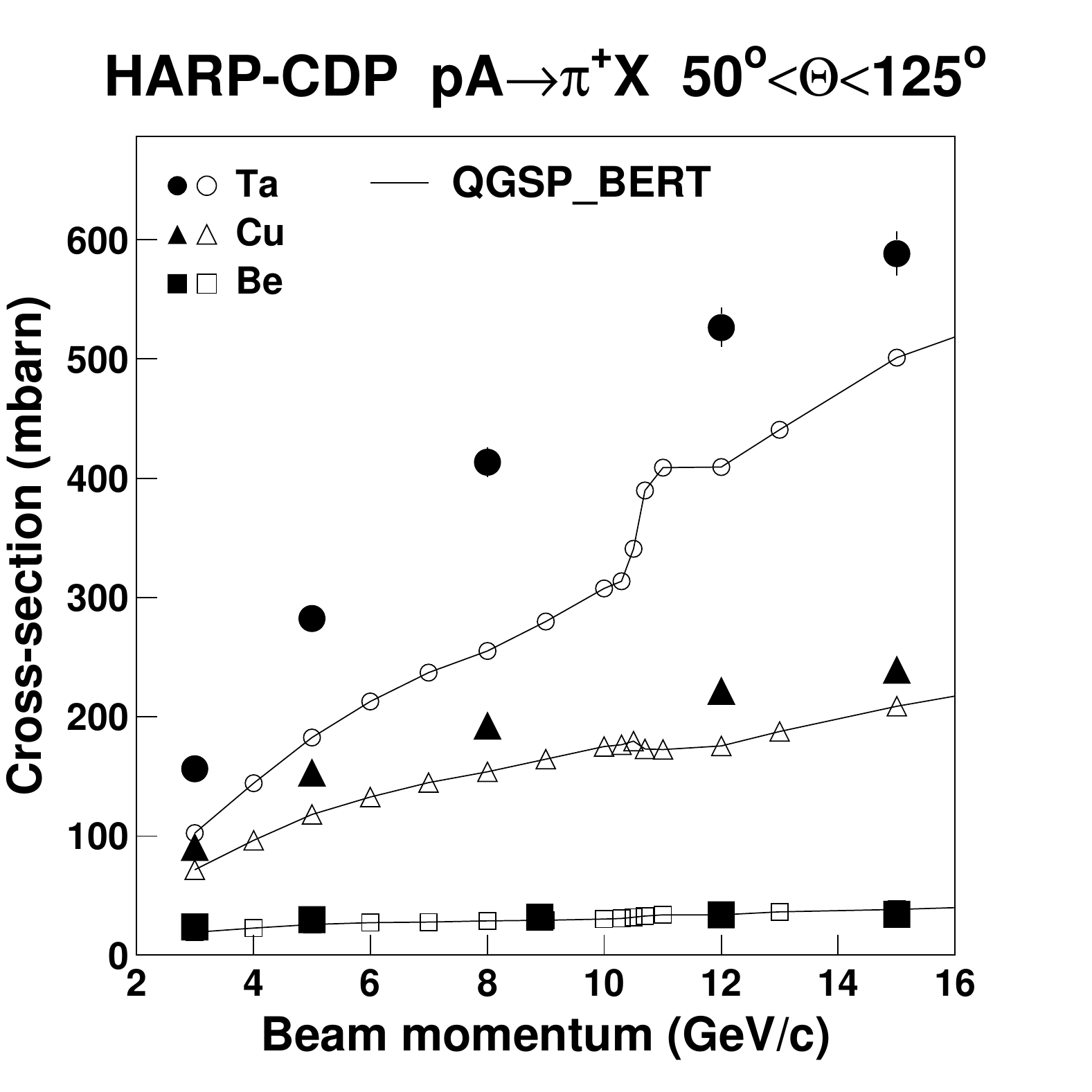} &
\includegraphics[width=0.45\textwidth]{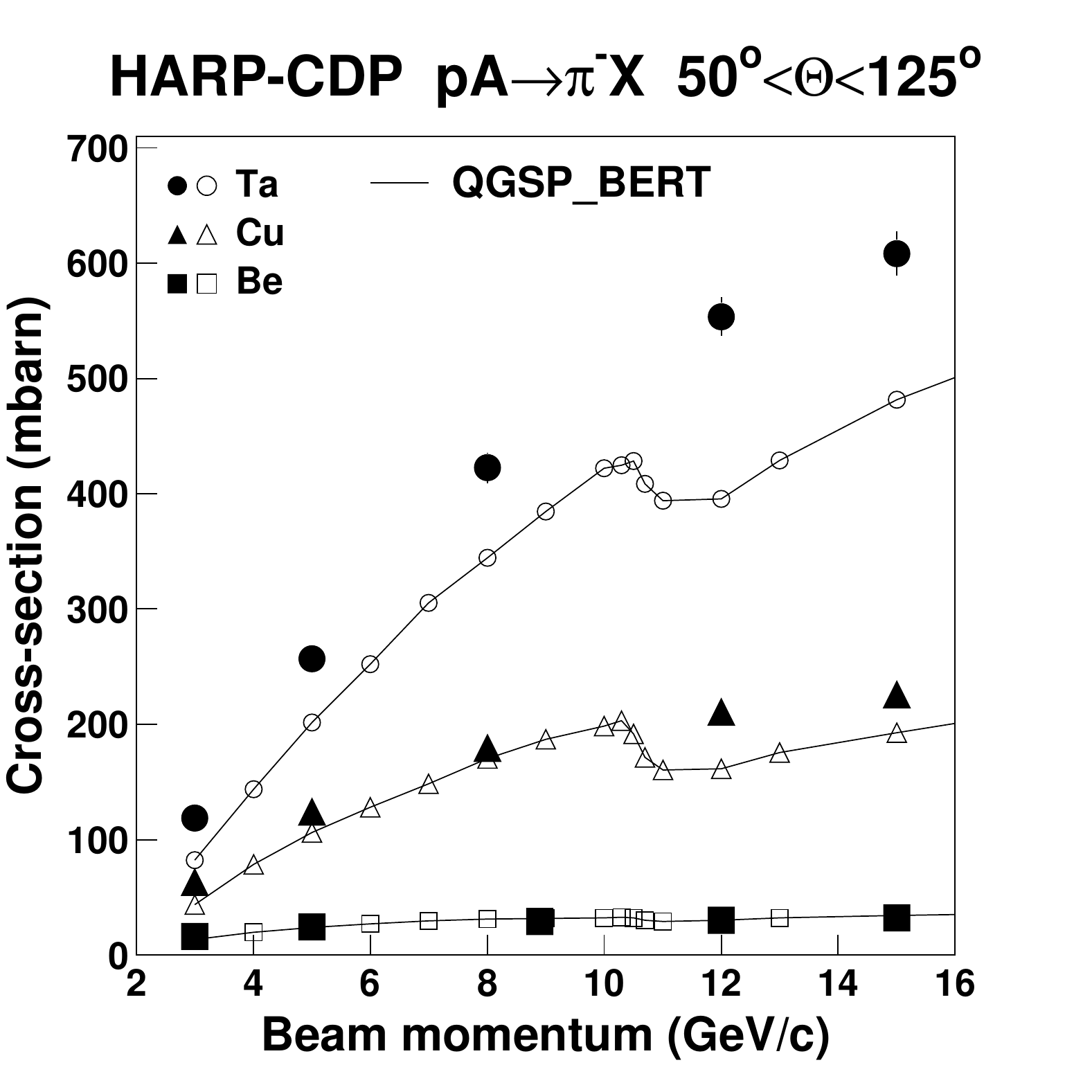} \\
\end{tabular}
\caption{Same as Fig.~\ref{FLUKAG4pionsbyprotons-int} but in the large-angle region.} 
\label{FLUKAG4pionsbyprotons-lar}
\end{center}
\end{figure}

Figures~\ref{FLUKAG4protonsbyprotons-int}--\ref{FLUKAG4pionsbyprotons-lar} show 
comparisons with FLUKA and Geant4 simulations 
of measured inclusive proton, $\pi^+$ and $\pi^-$ production cross-sections 
by protons on Be, Cu and Ta nuclei. For final-state protons, only comparisons 
in the intermediate-angle region are given (Fig.~\ref{FLUKAG4protonsbyprotons-int}). 
For final-state $\pi^+$'s and $\pi^-$'s, comparisons are given in the intermediate-angle 
(Fig.~\ref{FLUKAG4pionsbyprotons-int}) and large-angle regions 
(Fig.~\ref{FLUKAG4pionsbyprotons-lar}). 

\begin{figure}[h]
\begin{center}
\begin{tabular}{c}
\includegraphics[width=0.45\textwidth]{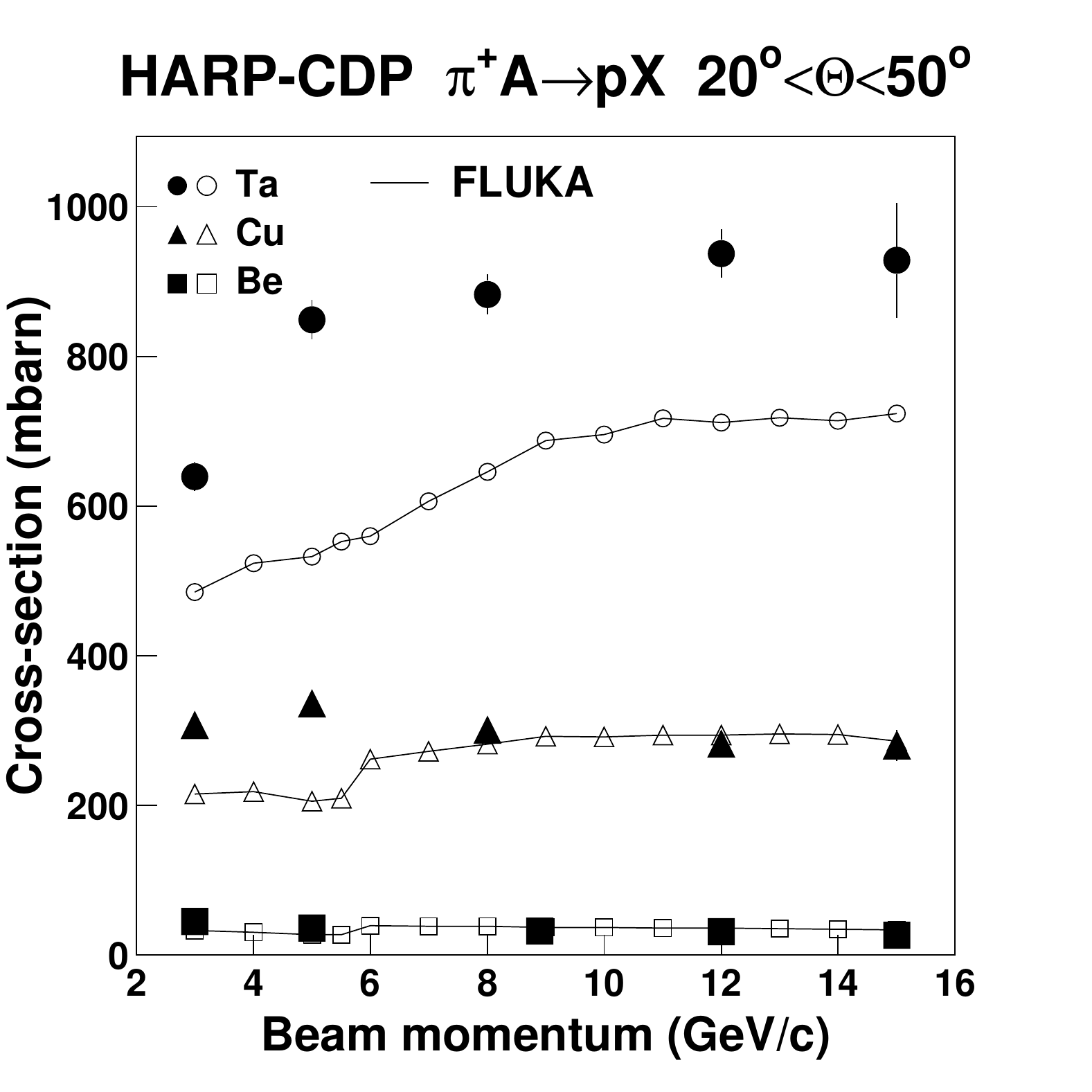} \\
\includegraphics[width=0.45\textwidth]{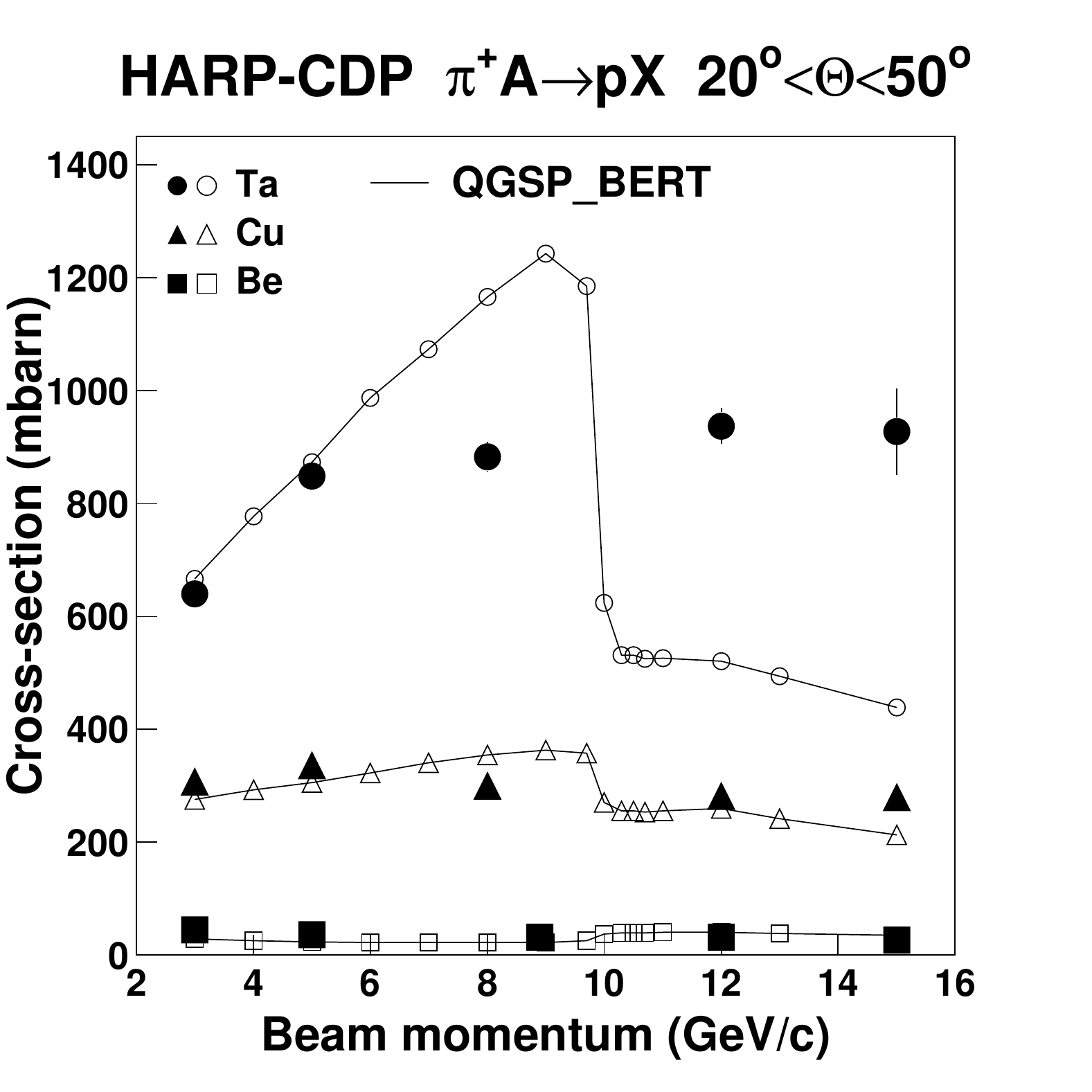} \\
\end{tabular}
\caption{Comparison of measured (black symbols) inclusive proton production
cross-sections by $\pi^+$'s on Be, Cu and Ta nuclei,  in the intermediate-angle region, as a
function of beam momentum, with FLUKA (upper panel) and Geant4 (lower panel) 
simulations (open symbols).} 
\label{FLUKAG4protonsbypiplus-int}
\end{center}
\end{figure}

\begin{figure}[h]
\begin{center}
\begin{tabular}{cc}
\includegraphics[width=0.45\textwidth]{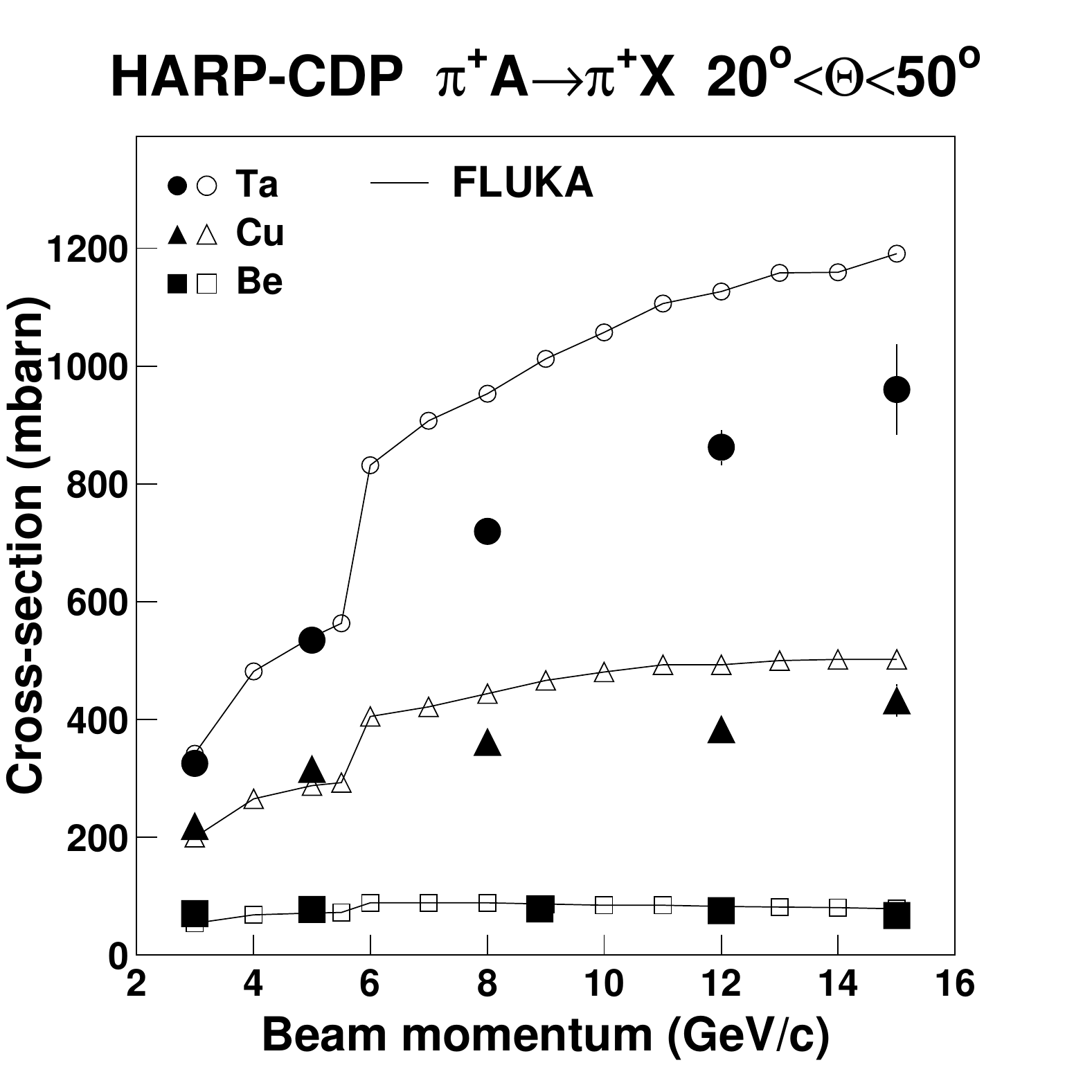} &
\includegraphics[width=0.45\textwidth]{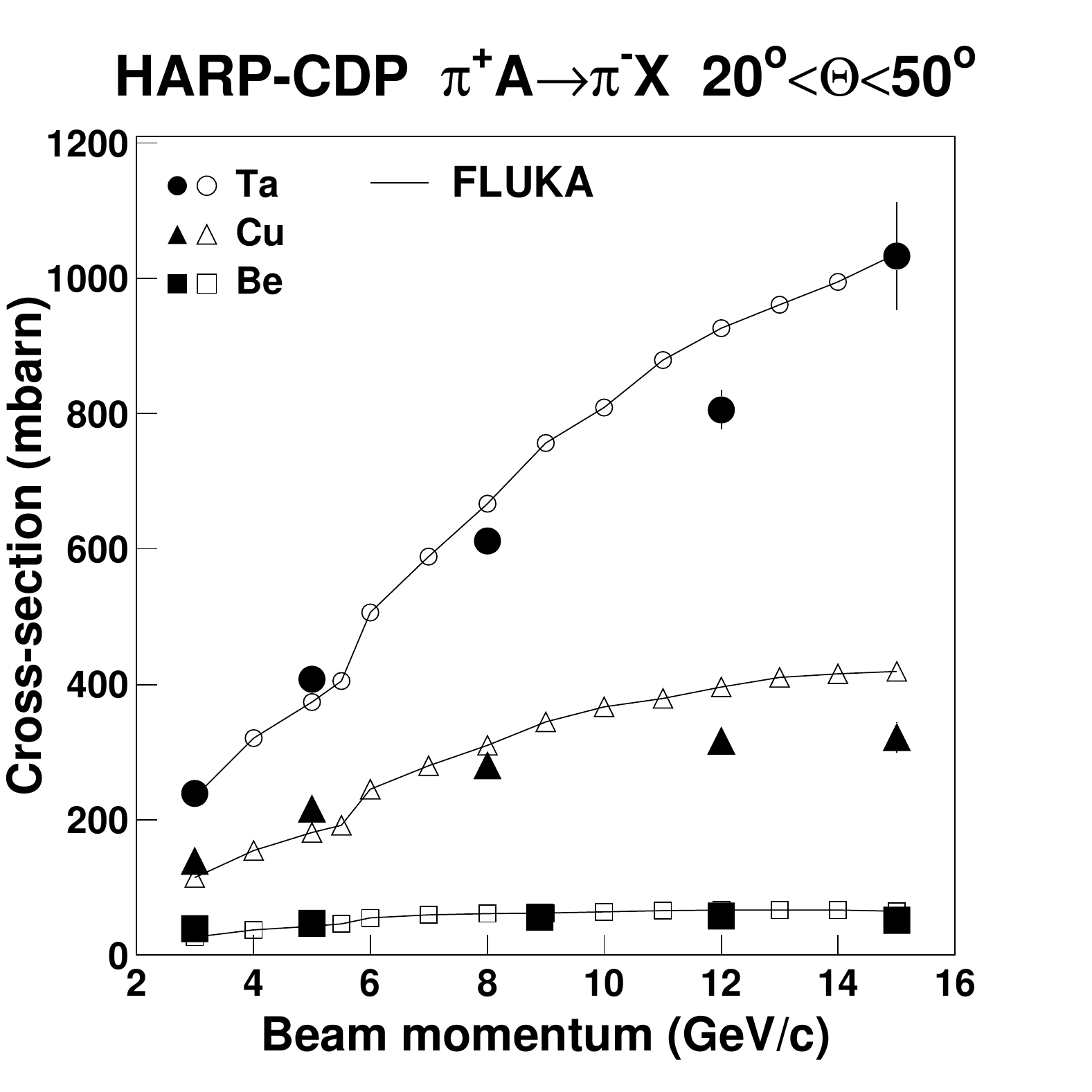} \\
\includegraphics[width=0.45\textwidth]{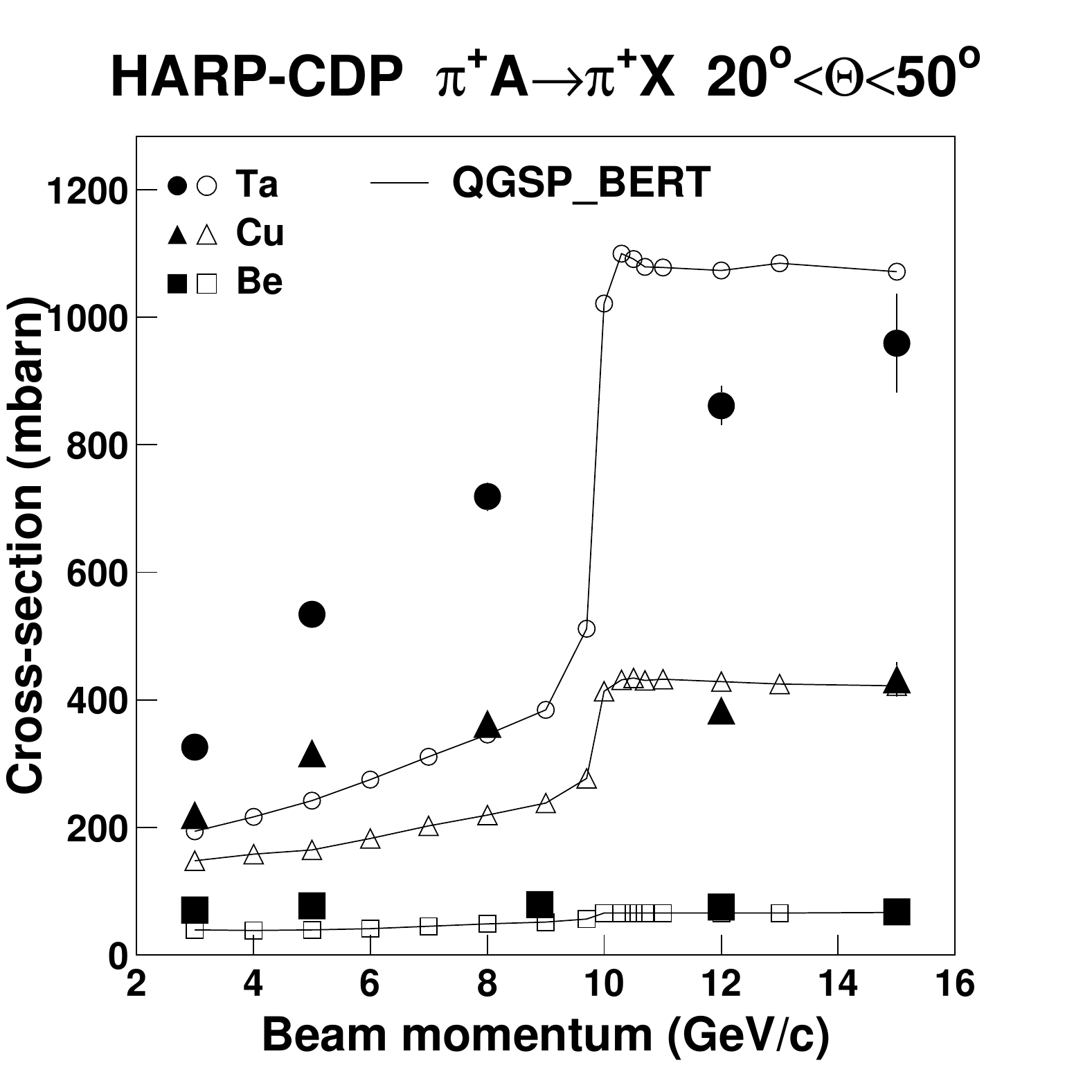} &
\includegraphics[width=0.45\textwidth]{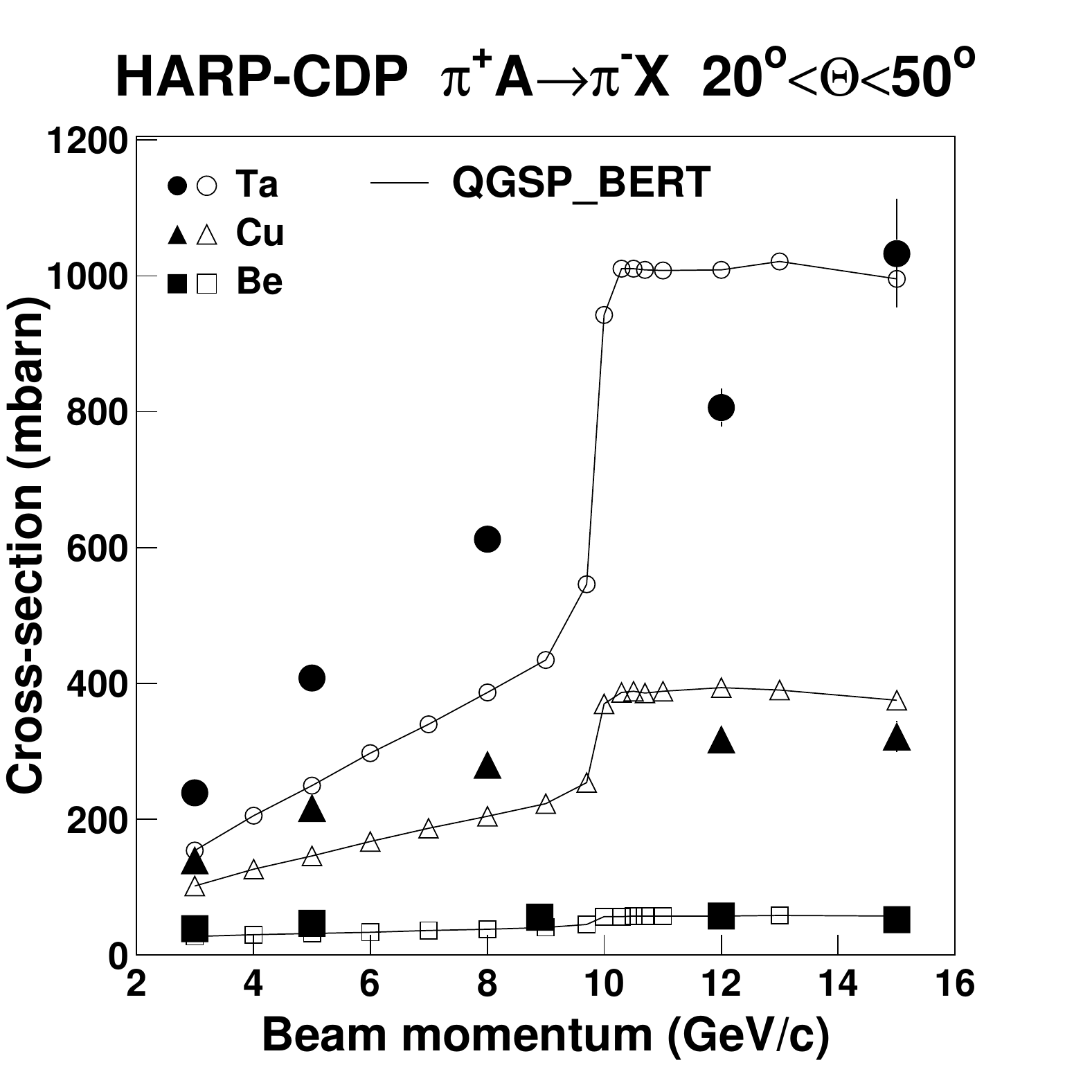} \\
\end{tabular}
\caption{Comparison of measured inclusive $\pi^+$ (left panels) and $\pi^-$  (right panels)production cross-sections
by $\pi^+$'s on Be, Cu and Ta nuclei (black symbols),  in the intermediate-angle region, 
with FLUKA and Geant4 simulations (open symbols).} 
\label{FLUKAG4pionsbypiplus-int}
\end{center}
\end{figure}

\begin{figure}[h]
\begin{center}
\begin{tabular}{cc}
\includegraphics[width=0.45\textwidth]{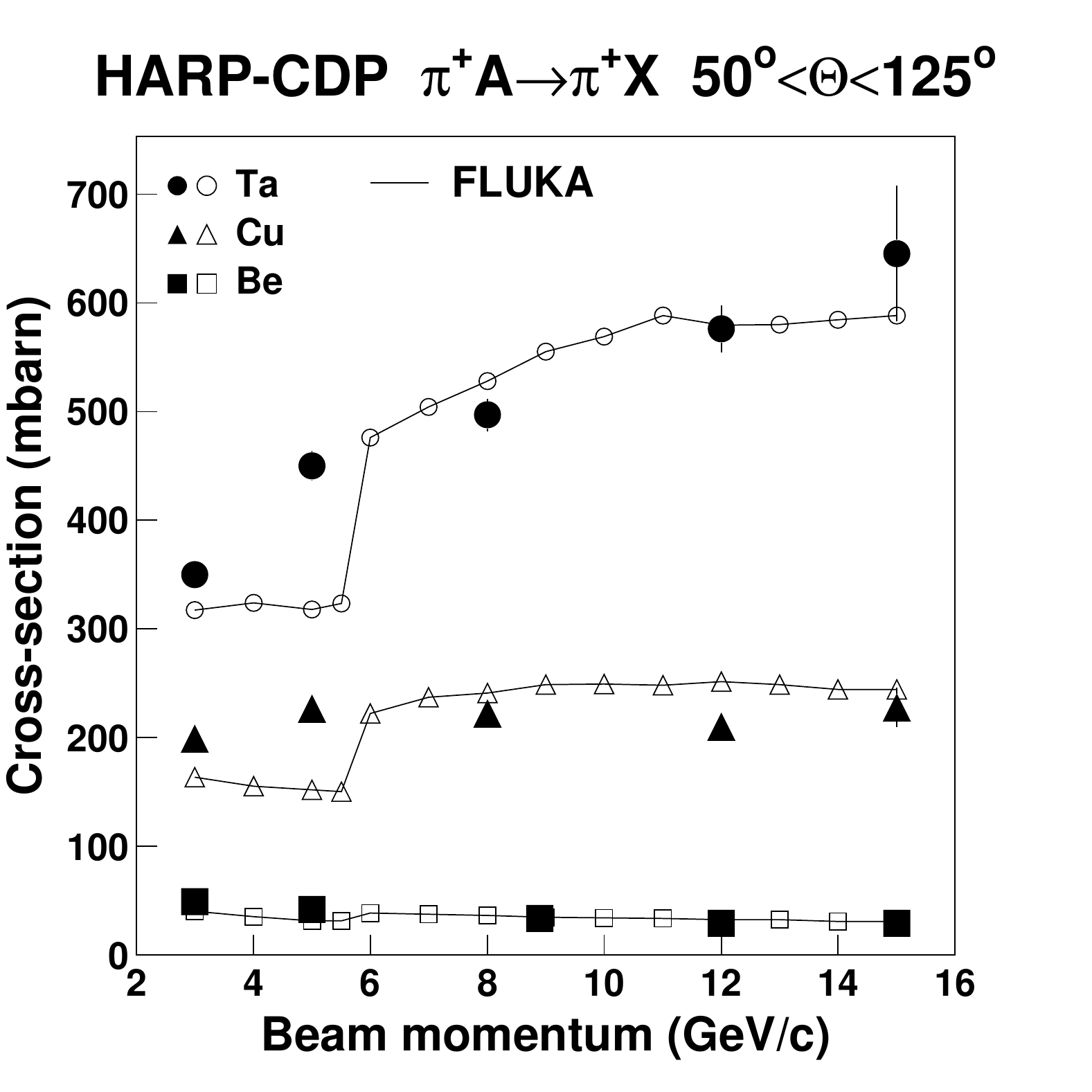} &
\includegraphics[width=0.45\textwidth]{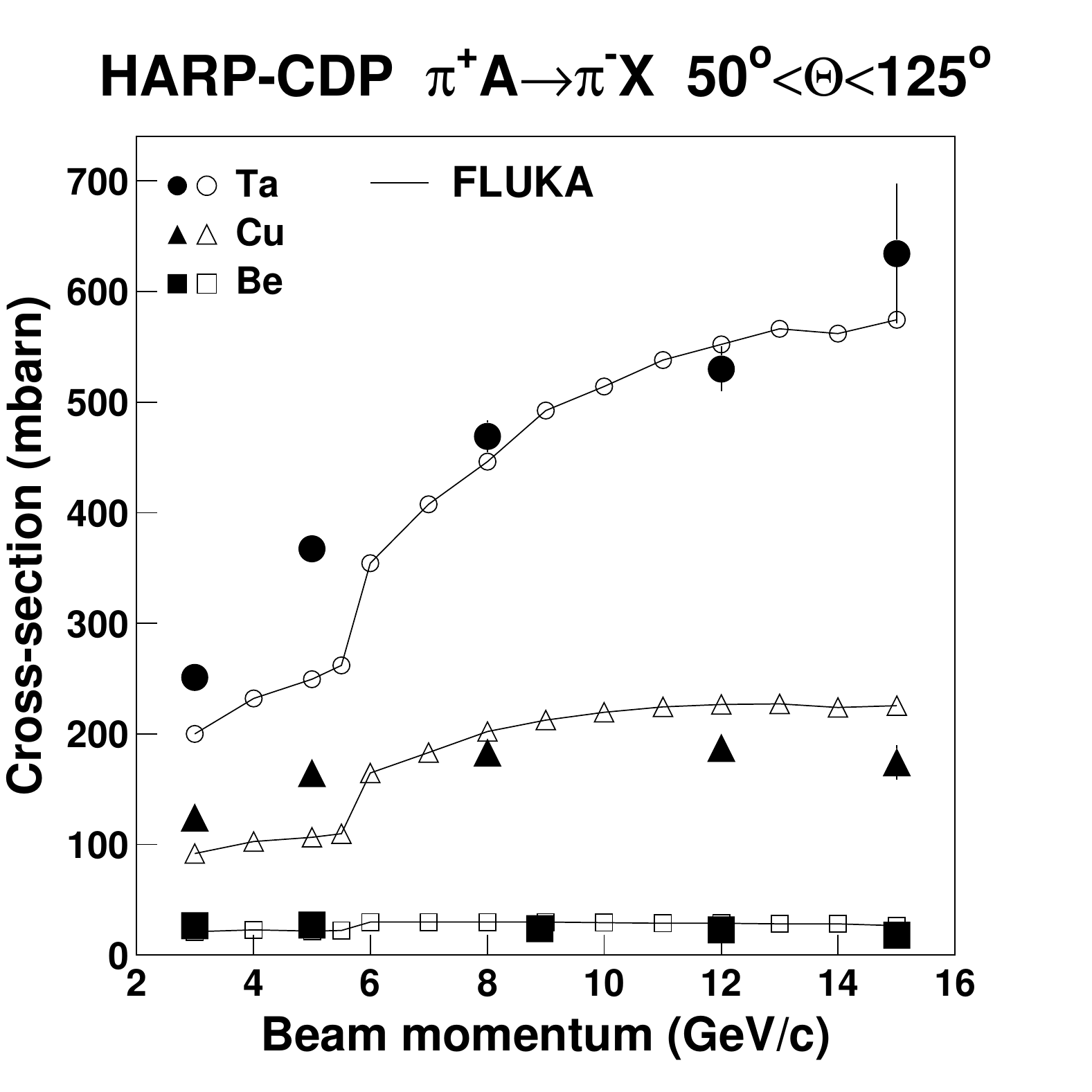} \\
\includegraphics[width=0.45\textwidth]{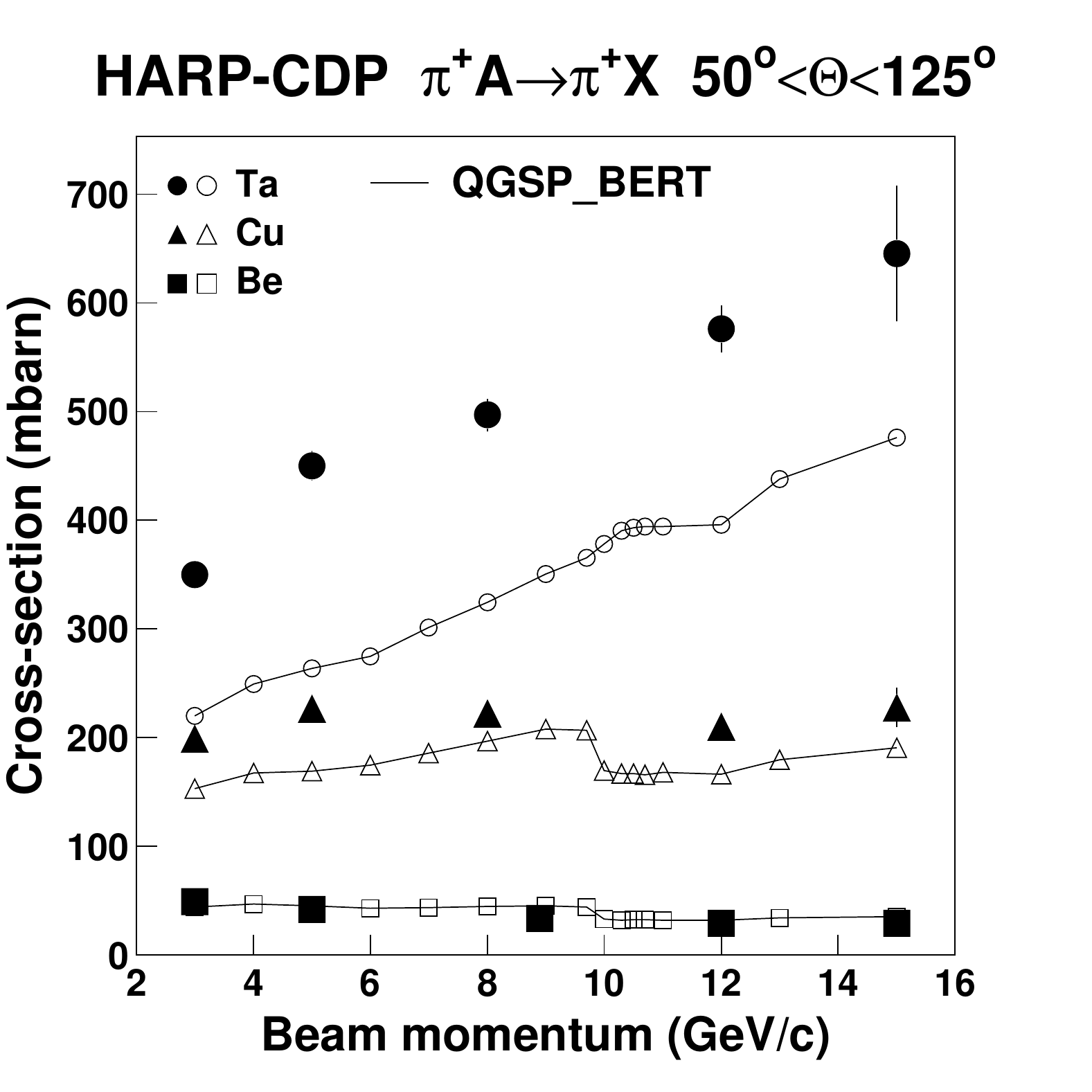} &
\includegraphics[width=0.45\textwidth]{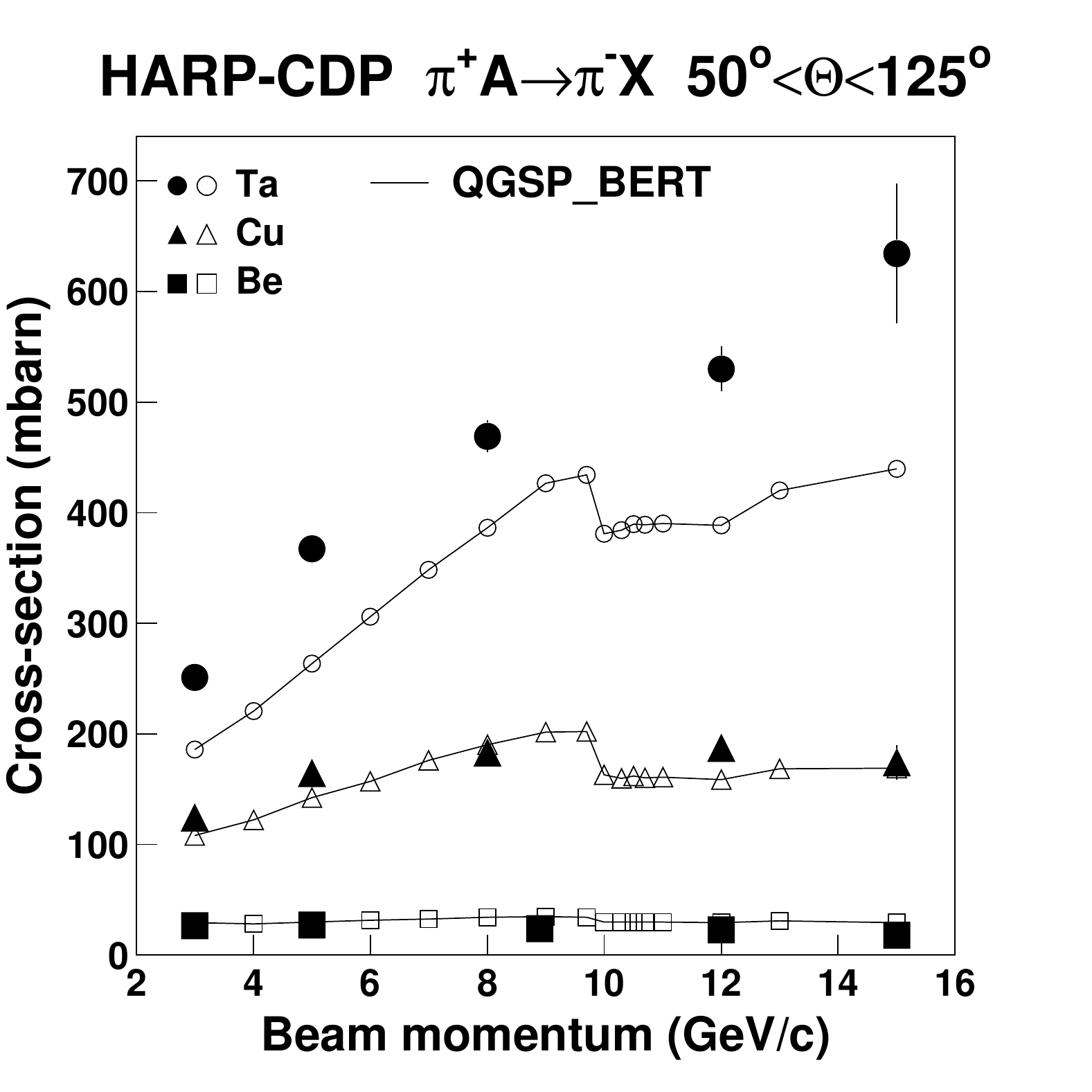} \\
\end{tabular}
\caption{Same as Fig.~\ref{FLUKAG4pionsbypiplus-int} but in the large-angle region.} 
\label{FLUKAG4pionsbypiplus-lar}
\end{center}
\end{figure}

\begin{figure}[h]
\begin{center}
\begin{tabular}{c}
\includegraphics[width=0.45\textwidth]{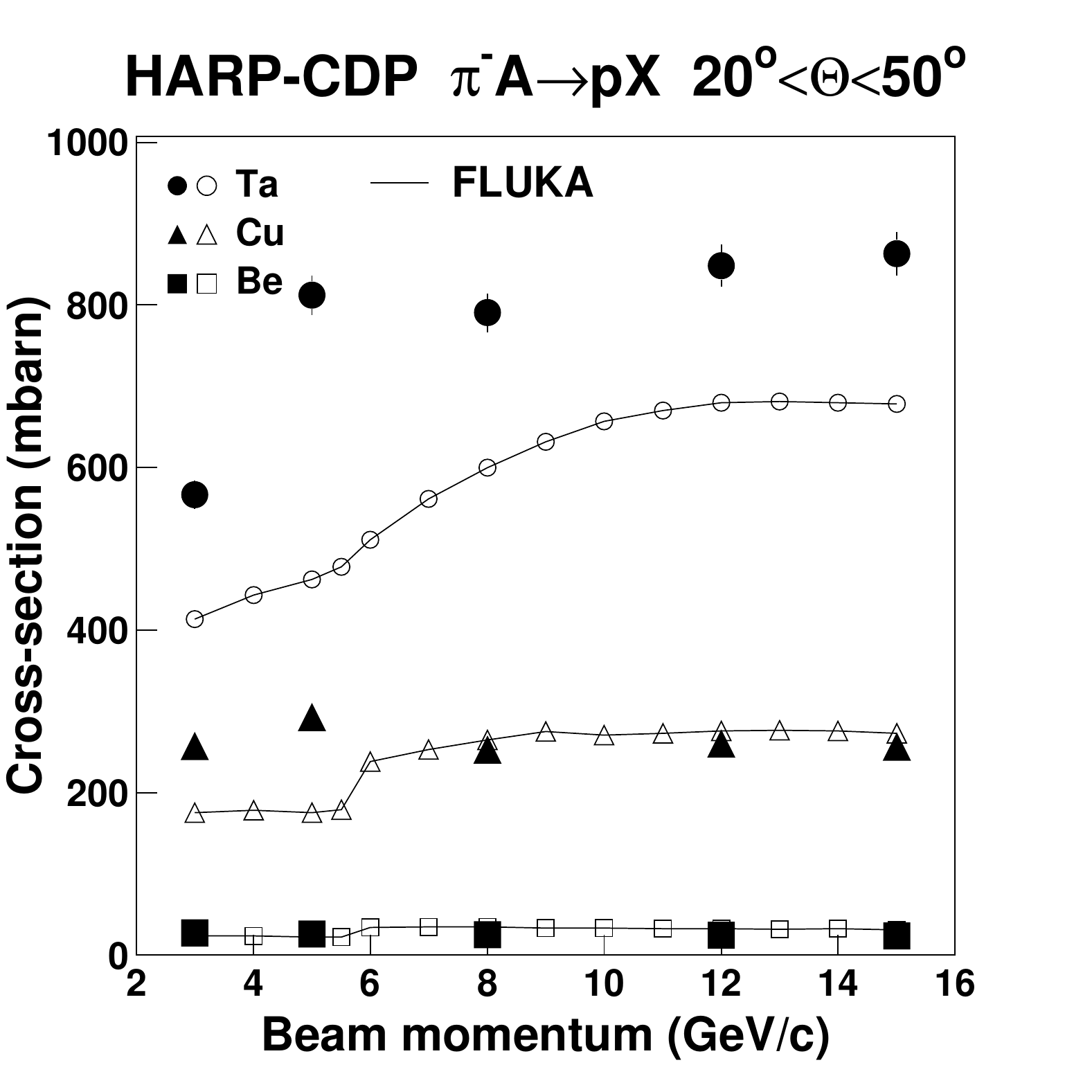} \\
\includegraphics[width=0.45\textwidth]{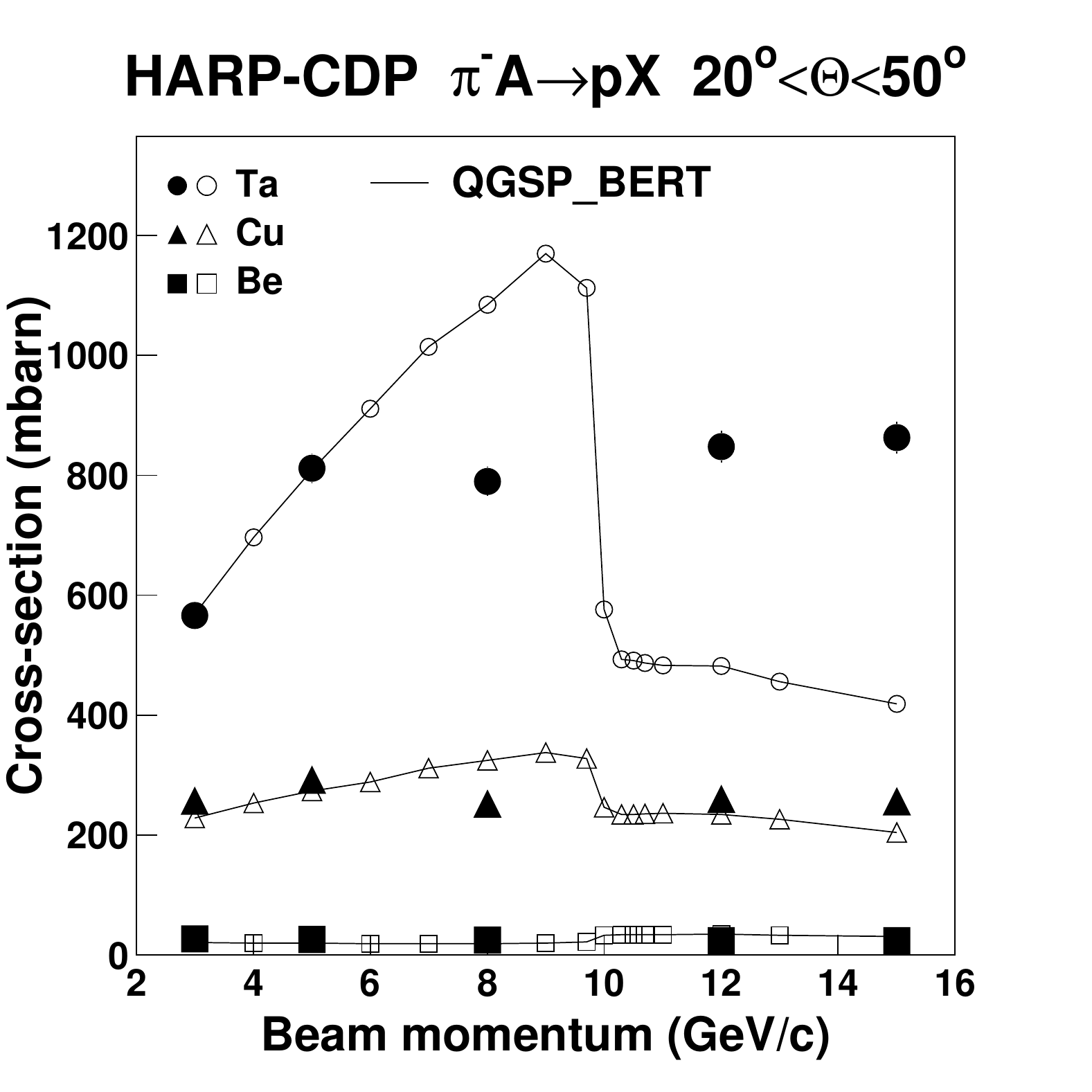} \\
\end{tabular}
\caption{Comparison of measured (black symbols) inclusive proton production
cross-sections by $\pi^-$'s on Be, Cu and Ta nuclei,  in the intermediate-angle region, as a
function of beam momentum, with FLUKA (upper panel) and Geant4 (lower panel) 
simulations (open symbols).} 
\label{FLUKAG4protonsbypiminus-int}
\end{center}
\end{figure}

\begin{figure}[h]
\begin{center}
\begin{tabular}{cc}
\includegraphics[width=0.45\textwidth]{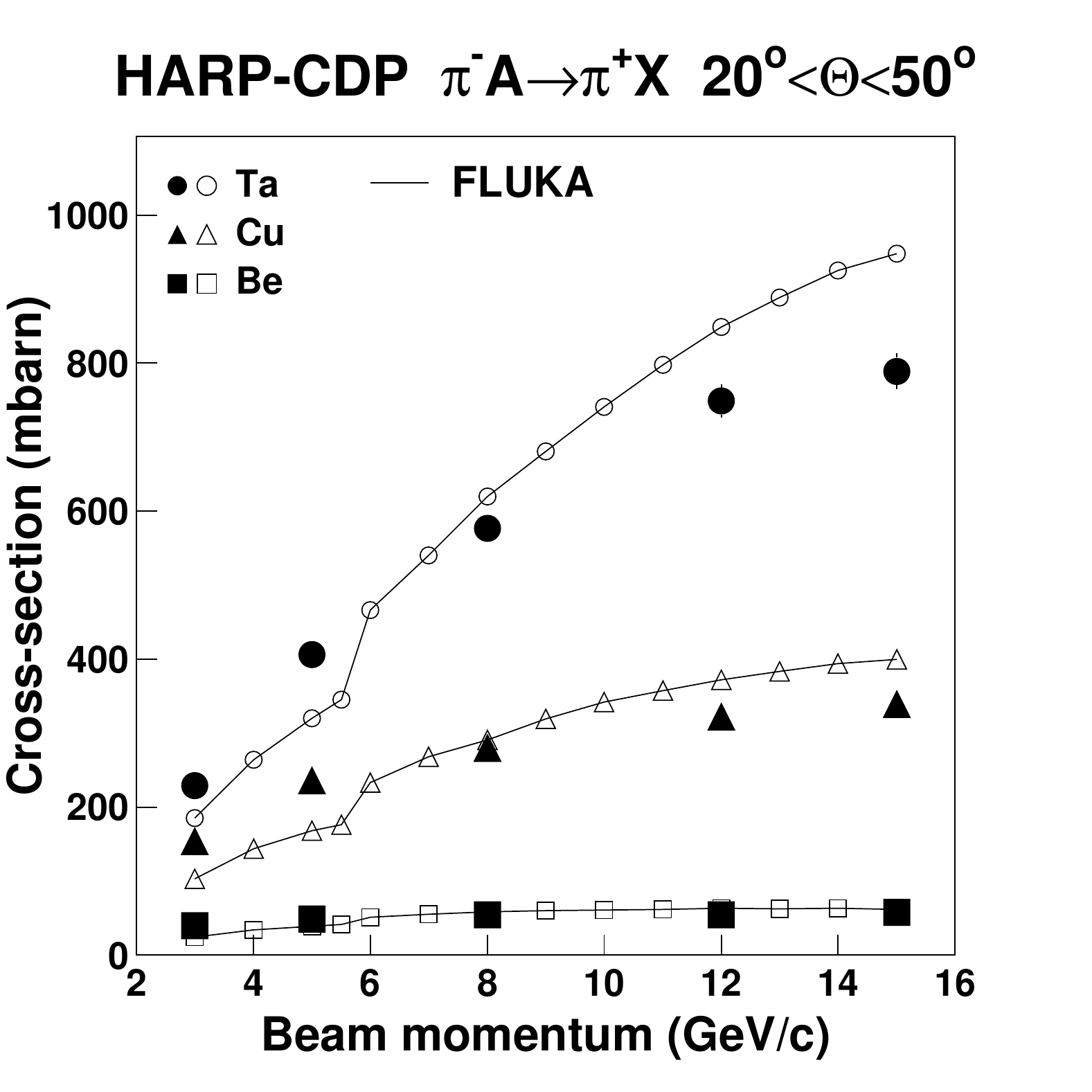} &
\includegraphics[width=0.45\textwidth]{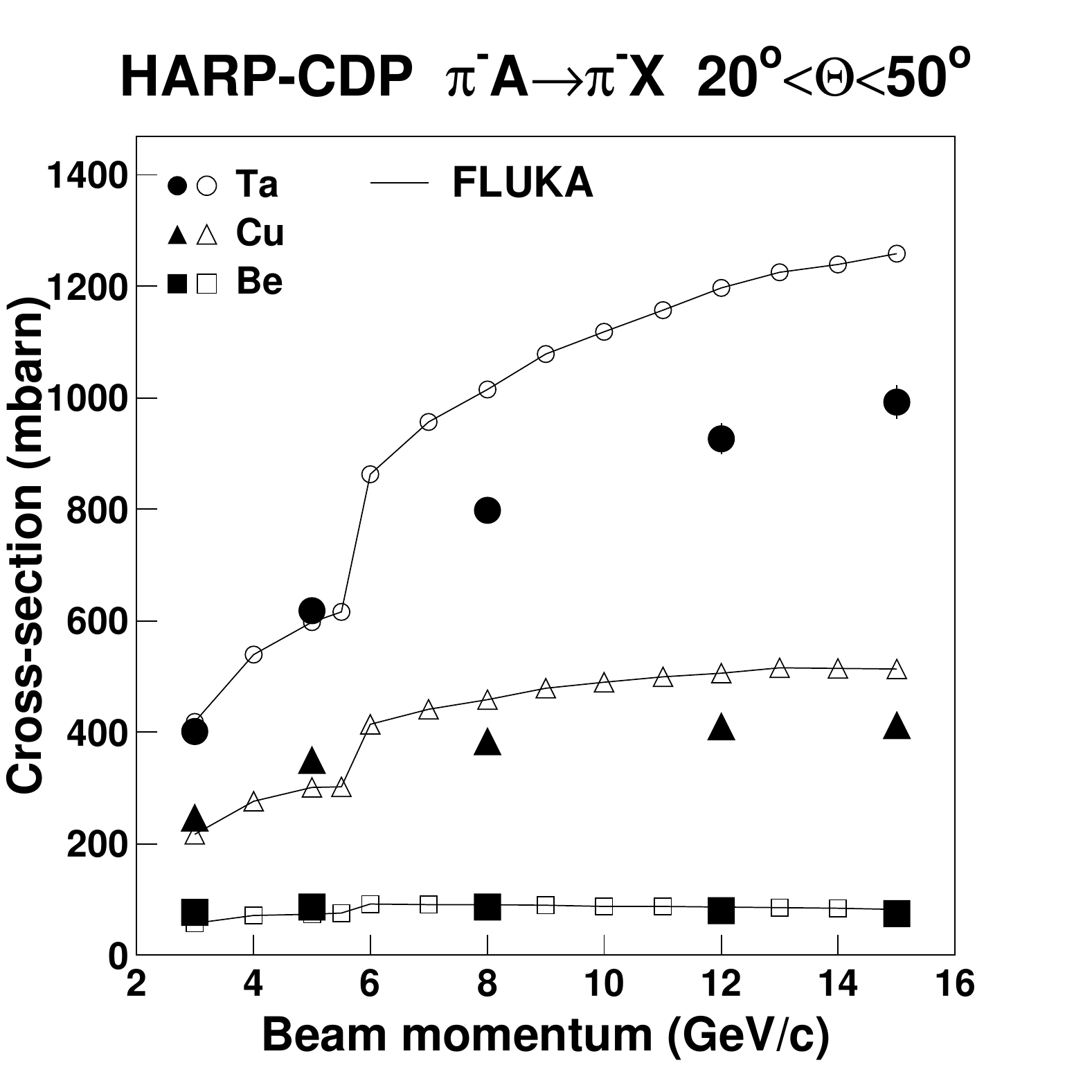} \\
\includegraphics[width=0.45\textwidth]{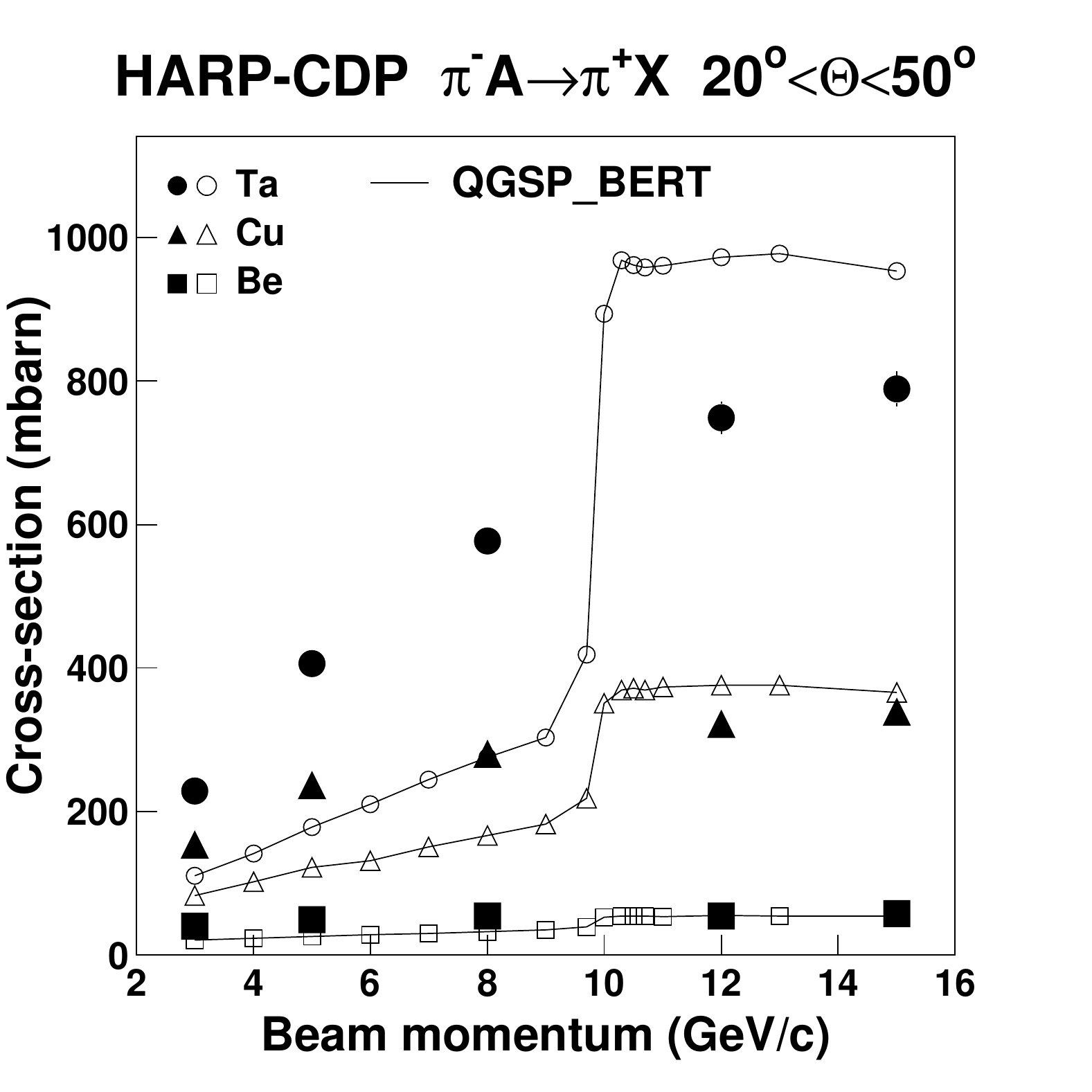} &
\includegraphics[width=0.45\textwidth]{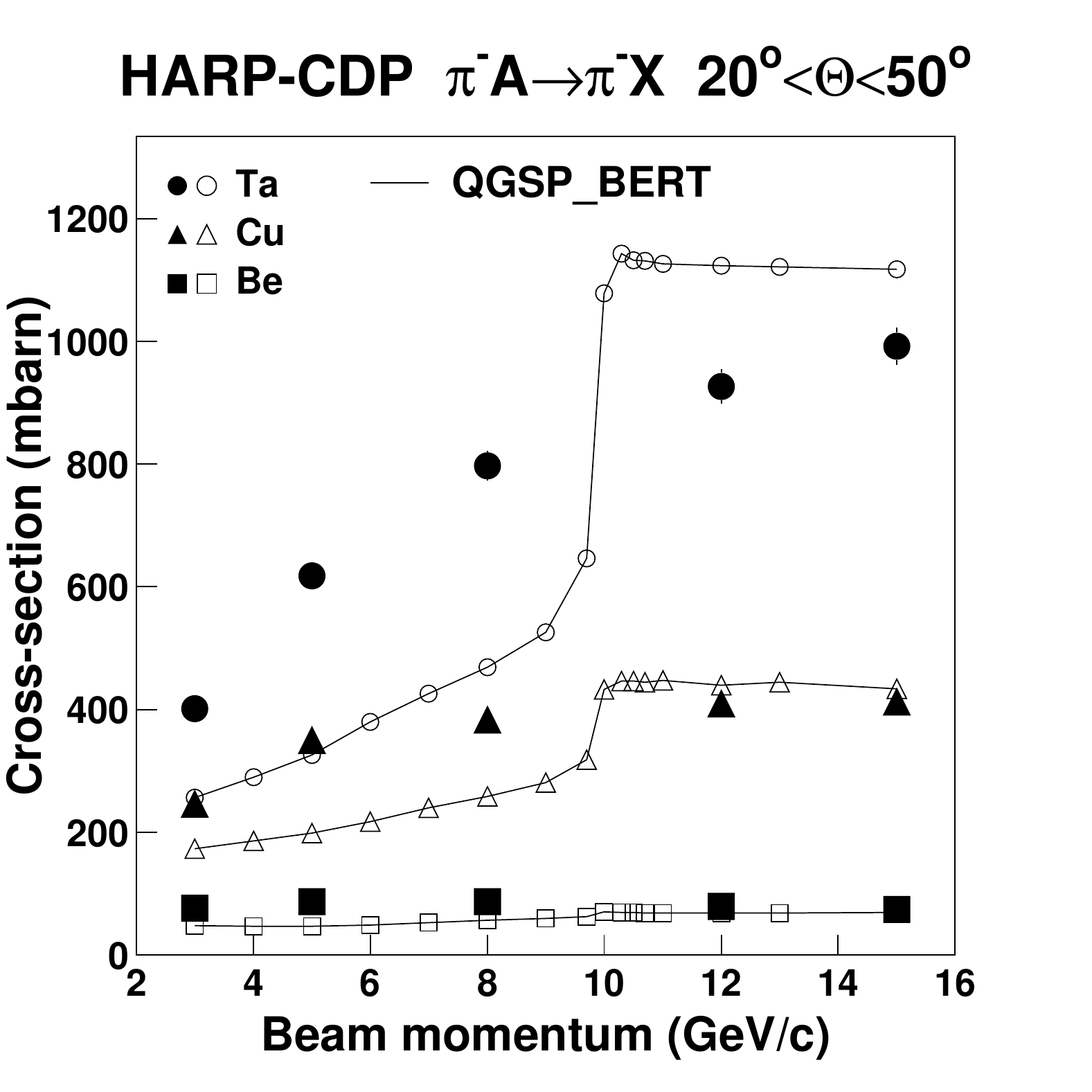} \\
\end{tabular}
\caption{Comparison of measured inclusive $\pi^+$ (left panels) and $\pi^-$  (right panels)production cross-sections
by $\pi^-$'s on Be, Cu and Ta nuclei (black symbols),  in the intermediate-angle region, 
with FLUKA and Geant4 simulations (open symbols).} 
\label{FLUKAG4pionsbypiminus-int}
\end{center}
\end{figure}

\begin{figure}[h]
\begin{center}
\begin{tabular}{cc}
\includegraphics[width=0.45\textwidth]{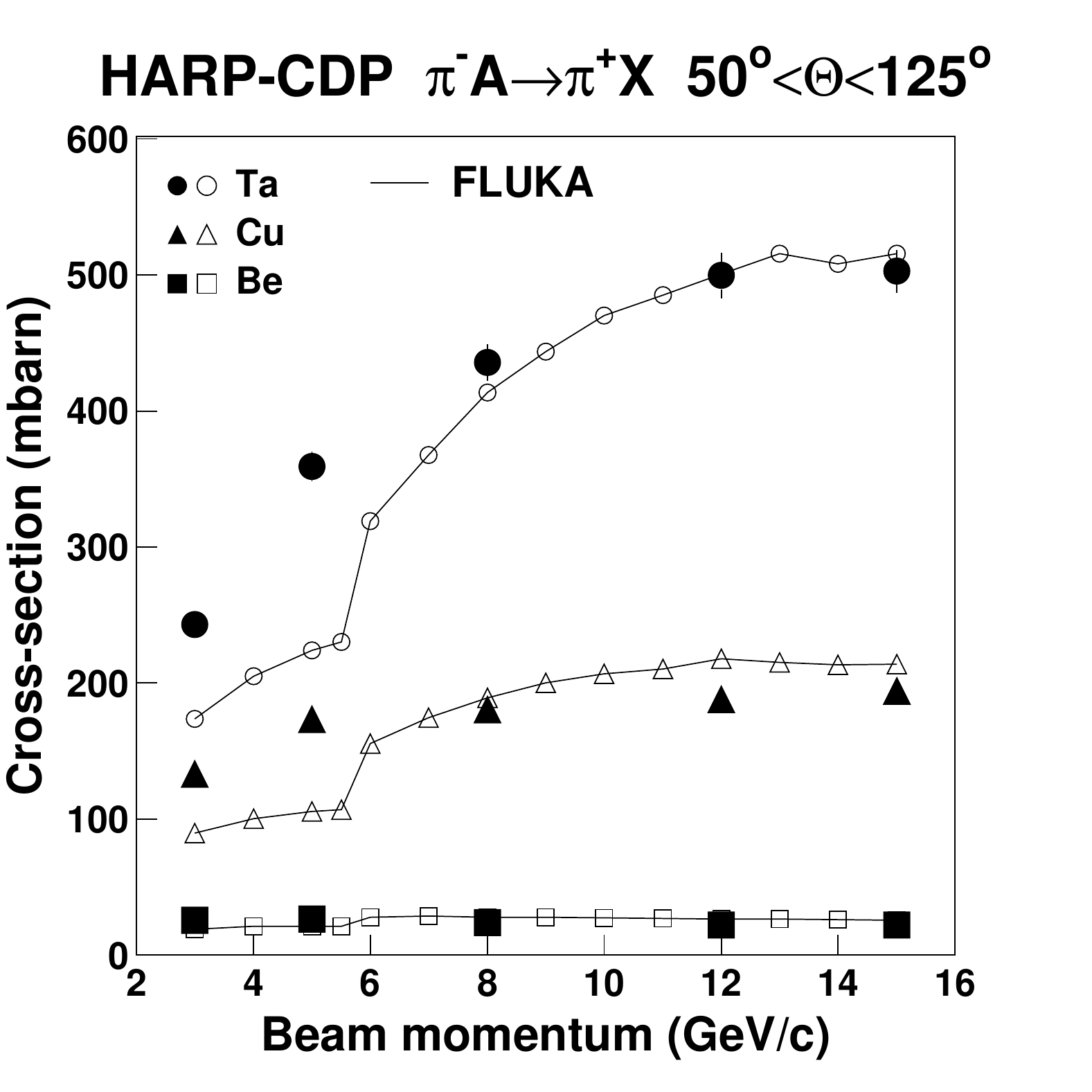} &
\includegraphics[width=0.45\textwidth]{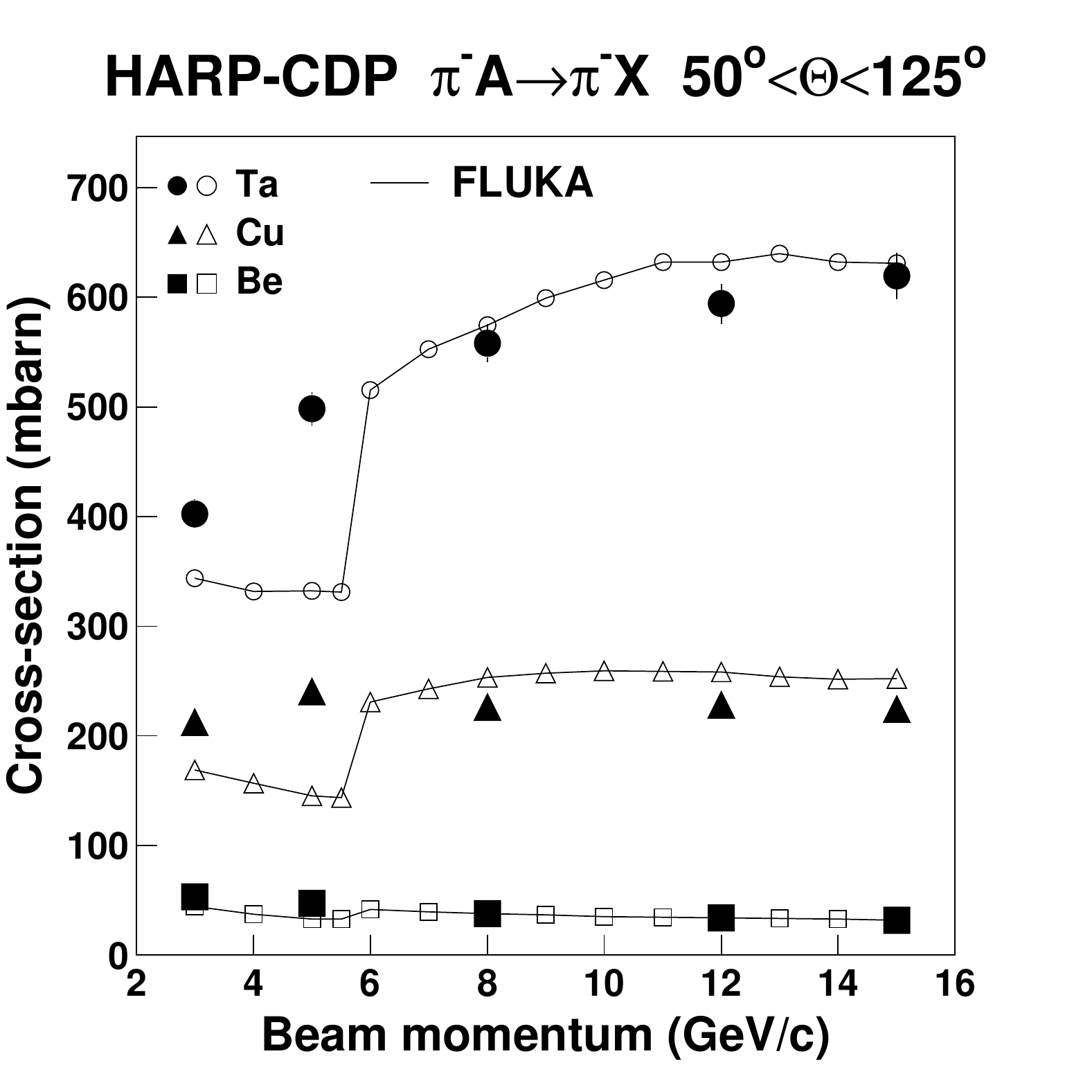} \\
\includegraphics[width=0.45\textwidth]{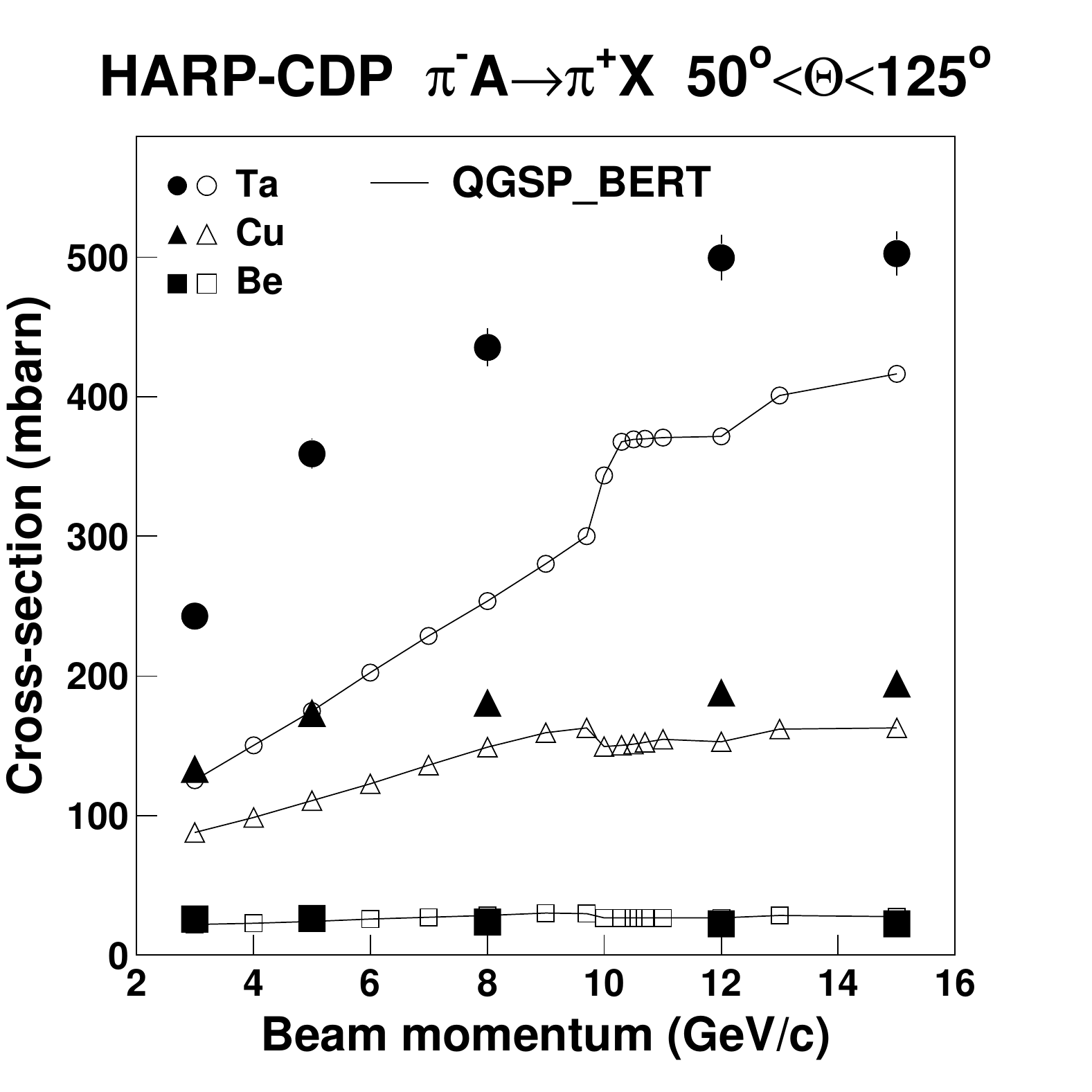} &
\includegraphics[width=0.45\textwidth]{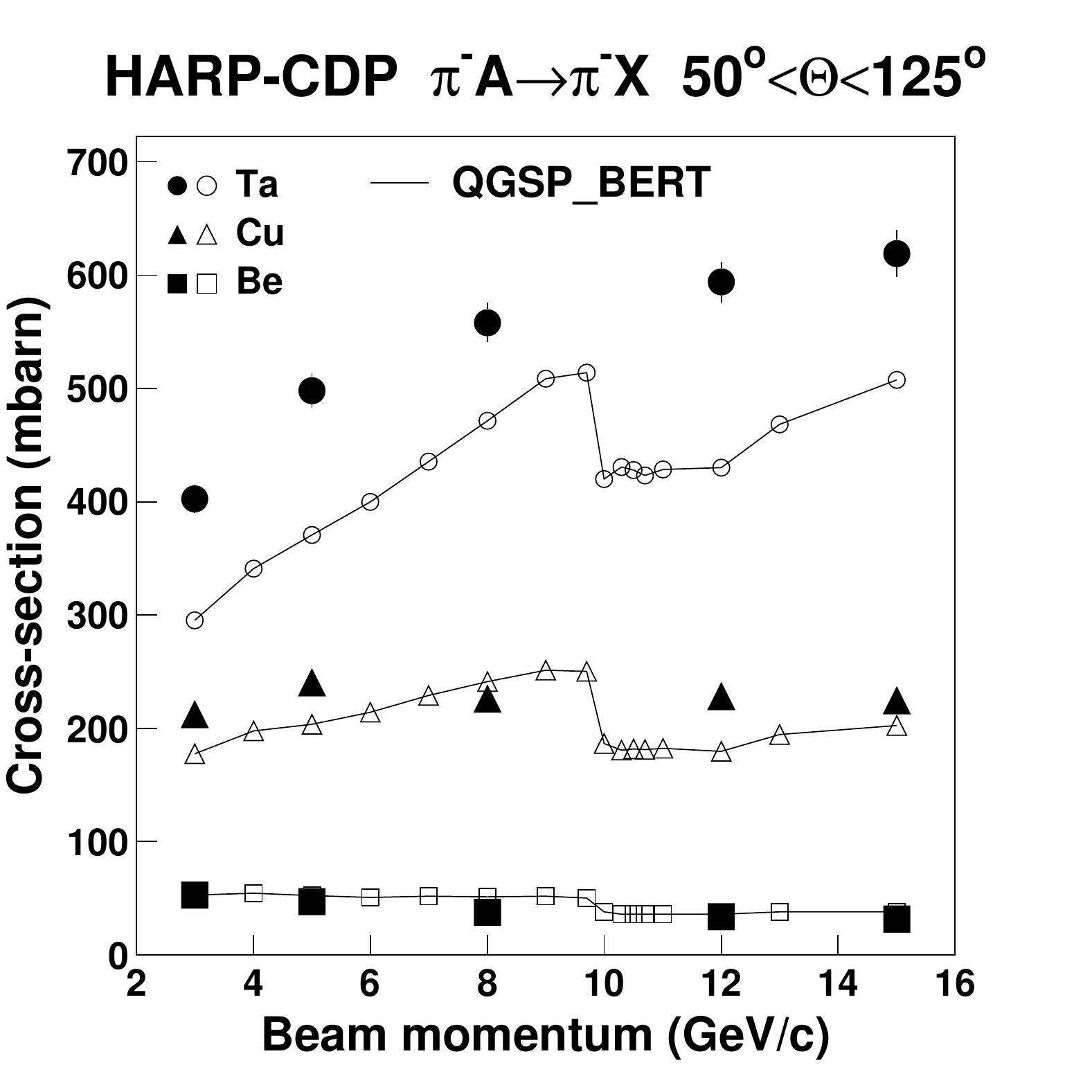} \\
\end{tabular}
\caption{Same as Fig.~\ref{FLUKAG4pionsbypiminus-int} but in the large-angle region.} 
\label{FLUKAG4pionsbypiminus-lar}
\end{center}
\end{figure}

While Figs.~\ref{FLUKAG4protonsbyprotons-int}--\ref{FLUKAG4pionsbyprotons-lar} 
show comparisons for proton beam particles, 
Figures~\ref{FLUKAG4protonsbypiplus-int}--\ref{FLUKAG4pionsbypiplus-lar} and 
Figs.~\ref{FLUKAG4protonsbypiminus-int}--\ref{FLUKAG4pionsbypiminus-lar} 
show the same comparisons for $\pi^+$ and $\pi^-$ beam particles, respectively.

\clearpage

\section{Effects from re-interactions of hadrons in nuclear matter}
\label{reinteractions}

Before our appraisal of the comparison of data with simulations
in Section~\ref{Appraisal}, it is useful to point to
generic features of final-state hadrons
from heavy nuclei.

\begin{figure}[htp]
\begin{center}
\includegraphics[width=0.45\textwidth]{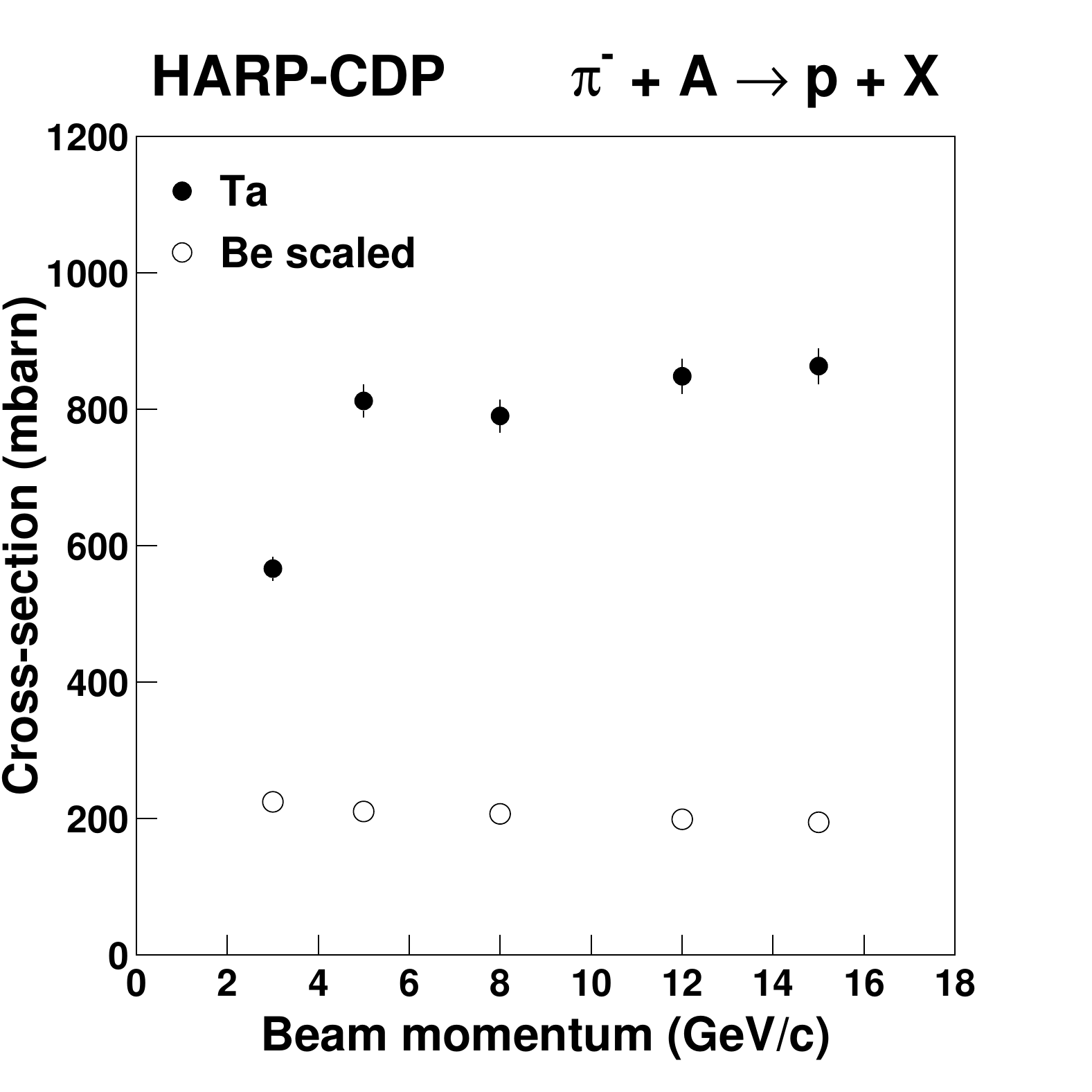} 
\caption{Inclusive cross-sections of proton-production with polar angle 
$20^\circ < \theta < 50^\circ$ by incoming $\pi^-$ beam particles 
with momenta in the range 3--15~GeV/{\it c}, for beryllium and and tantalum nuclei;
The tantalum cross-sections are those measured, while the beryllium
cross-sections are scaled with $(A_{\rm Ta}/A_{\rm Be})^{0.7}$. }
\label{scaledcrosssections}
\end{center}
\end{figure}

Fig.~\ref{scaledcrosssections} shows
inclusive cross-sections of proton-production with polar angle 
$20^\circ < \theta < 50^\circ$ by incoming $\pi^-$ beam particles 
with momenta in the range 3--15~GeV/{\it c}, for beryllium and and tantalum nuclei.
The tantalum cross-sections are shown as measured, while the beryllium
cross-sections are scaled with $(A_{\rm Ta}/A_{\rm Be})^{0.7}$. The rationale is 
that the latter cross-sections are the ones on a hypothetical 
tantalum nucleus that has the same small
re-interaction probability of secondaries as a beryllium nucleus. 
Fig.~\ref{scaledcrosssections} tells 
that the measured inclusive cross-section of tantalum is still four times
larger than the scaled beryllium cross-section. Therefore, the bulk of
final-state protons stem from re-interactions of secondaries in the nuclear
matter of the tantalum nucleus---which after all has a diameter that is
equivalent to several
nuclear interaction lengths.

In Fig.~\ref{scaledcrosssections}, it also appears that the inclusive cross-section of proton
production on tantalum nuclei at 3~GeV/{\it c} beam momentum 
does not follow the trend that is suggested by the data points at higher beam momentum. Indeed, a similar but smaller reduction of cross-section can be seen in the proton production by incoming beam $\pi^+$'s (see Fig.~\ref{FLUKAG4protonsbypiplus-int}) and an even smaller reduction also for incoming beam protons 
(see Fig.~\ref{FLUKAG4protonsbyprotons-int}). No such reduction is seen for the production of
secondary $\pi^+$'s and $\pi^-$'s. We conjecture that this reduction reflects the
absorption of low-momentum final-state protons in the nuclear matter of the
target nucleus.

\begin{figure}[htp]
\begin{center}
\begin{tabular}{c}
\includegraphics[width=0.45\textwidth]{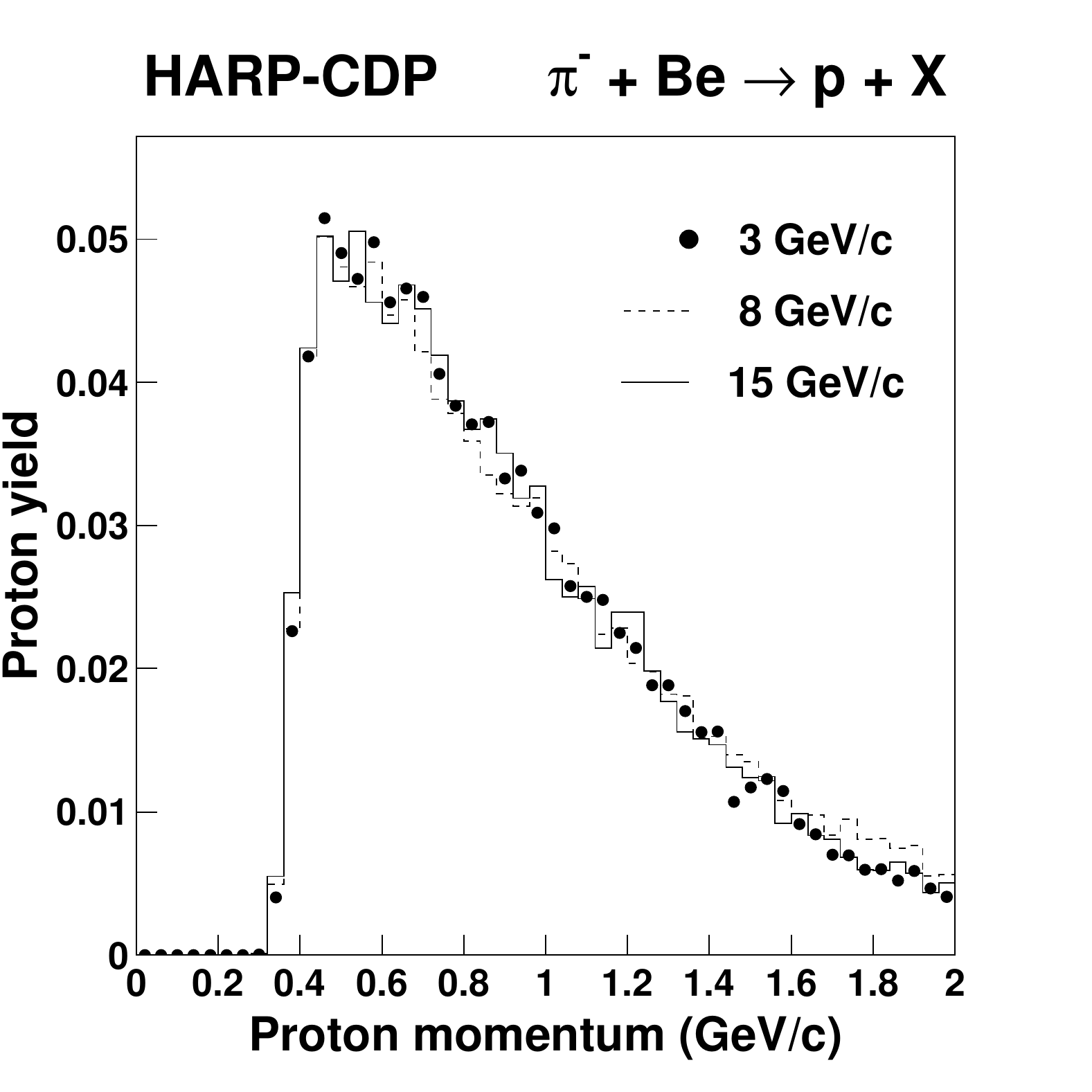} \\
\includegraphics[width=0.45\textwidth]{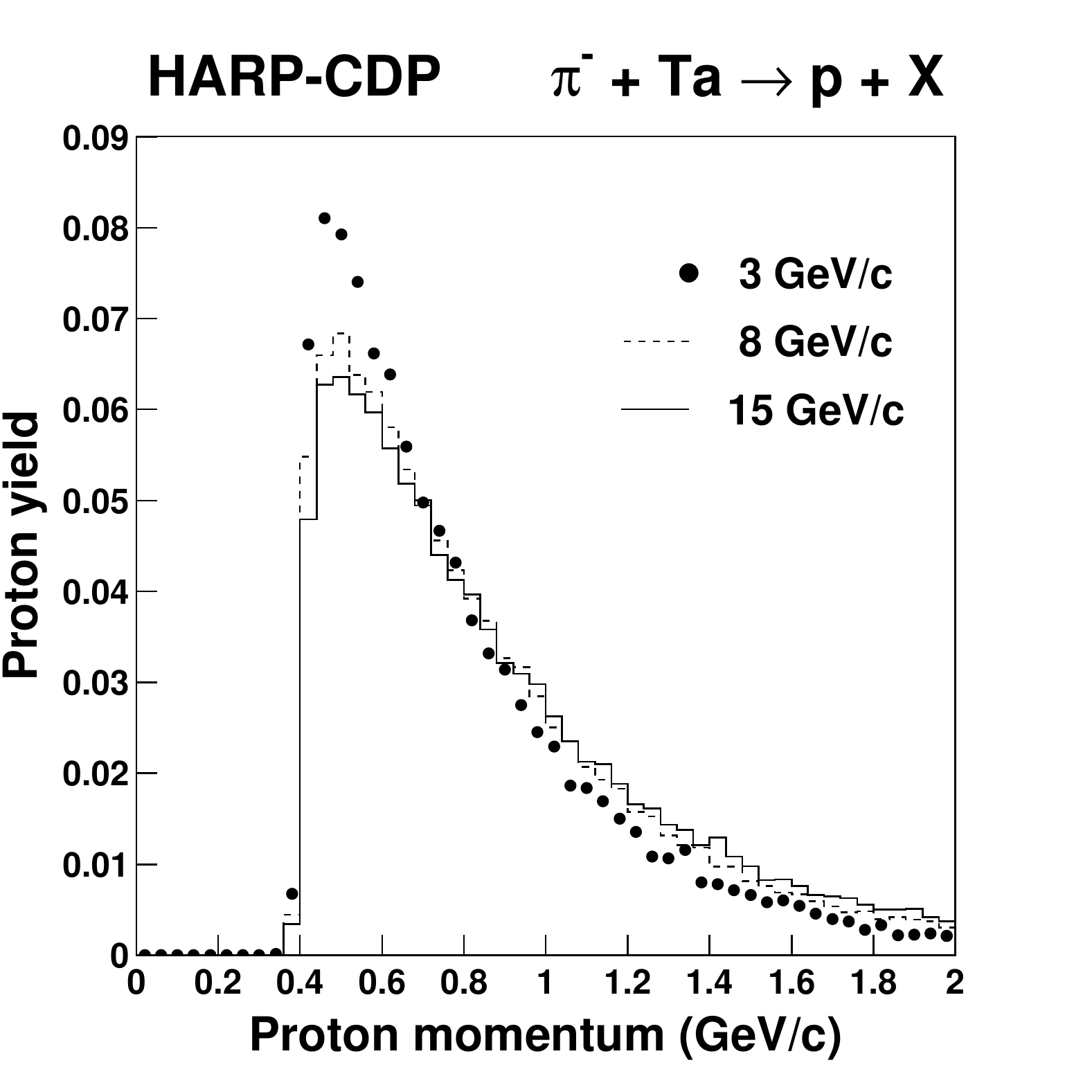} \\
\end{tabular}
\caption{Comparison of total-momentum distributions, normalized to unit area, 
of final-state protons with
polar angle $20^\circ  < \theta < 50^\circ$, for $\pi^-$ beam particle momenta of
3 GeV/{\it c} (black points), 8 GeV/{\it c} (dotted histogram)  and 15 GeV/{\it c} (full histogram), 
for beryllium (upper panel) and for tantalum (lower panel) target nuclei.} 
\label{pdistributions}
\end{center}
\end{figure}

With a view to corroborating this conjecture, Fig.~\ref{pdistributions} 
shows the measured total-momentum 
distributions of final-state protons, in the polar-angle range $20^\circ  < \theta < 50^\circ$, for incoming $\pi^-$ beam particles. Since the emphasis is on shape comparison, all distributions are normalized to unit area. The upper 
panel shows the comparison between beam momenta of 3, 8 and 15~GeV/{\it c} beam momentum for beryllium target nuclei, while the lower panel shows the same for tantalum target 
nuclei. The difference in the shape of the total-momentum distributions of final-state protons between Be and Ta, and for Ta between different $\pi^-$ beam particle momenta, is apparent. The heavier the nucleus and the lower the beam momentum, the more the spectrum is shifted toward
smaller momentum values. 
Because of the strong rise of the proton--nucleon cross-section for proton momenta below 0.8~GeV/{\it c}, many final-state protons are absorbed inside the target nucleus which causes a cross-section reduction which affects among our beam momenta  strongest
the final-state protons at 3~GeV/{\it c} beam momentum. The reduction is largest for $\pi^-$ beam particles because they are least efficient to produce high-momentum final-state protons that escape the nucleus. The reduction is not seen for final-state pions because the pion--nucleon cross-section permits at low momenta for pions an easier escape from the nucleus than for protons.


\section{Appraisal of the differences between data and simulations}
\label{Appraisal}

\subsection{FLUKA versus data}
\label{FLUKA_discussion}

In our parameter range of the comparison of data with FLUKA simulations,
the latter agree with data within some 30\%.

The most unsatisfactory feature in the FLUKA simulation is the discontinuity
around 5~GeV/{\it c} beam momentum which is rather persistent 
in all combinations of beam particle and final-state particle.

\subsection{Geant4 versus data}
\label{Geant4_discussion}

In our parameter range of the comparison of data with Geant4 QGSP\_BERT
simulations, the latter agree with data within a factor of about two.

An unacceptable feature of the Geant4 QGSP\_BERT simulation is the strong discontinuity
around 10~GeV/{\it c} beam momentum which is also rather persistent
in all combinations of beam particle and final-state particle.

\begin{figure}[h]
\begin{center}
\includegraphics[width=0.45\textwidth]{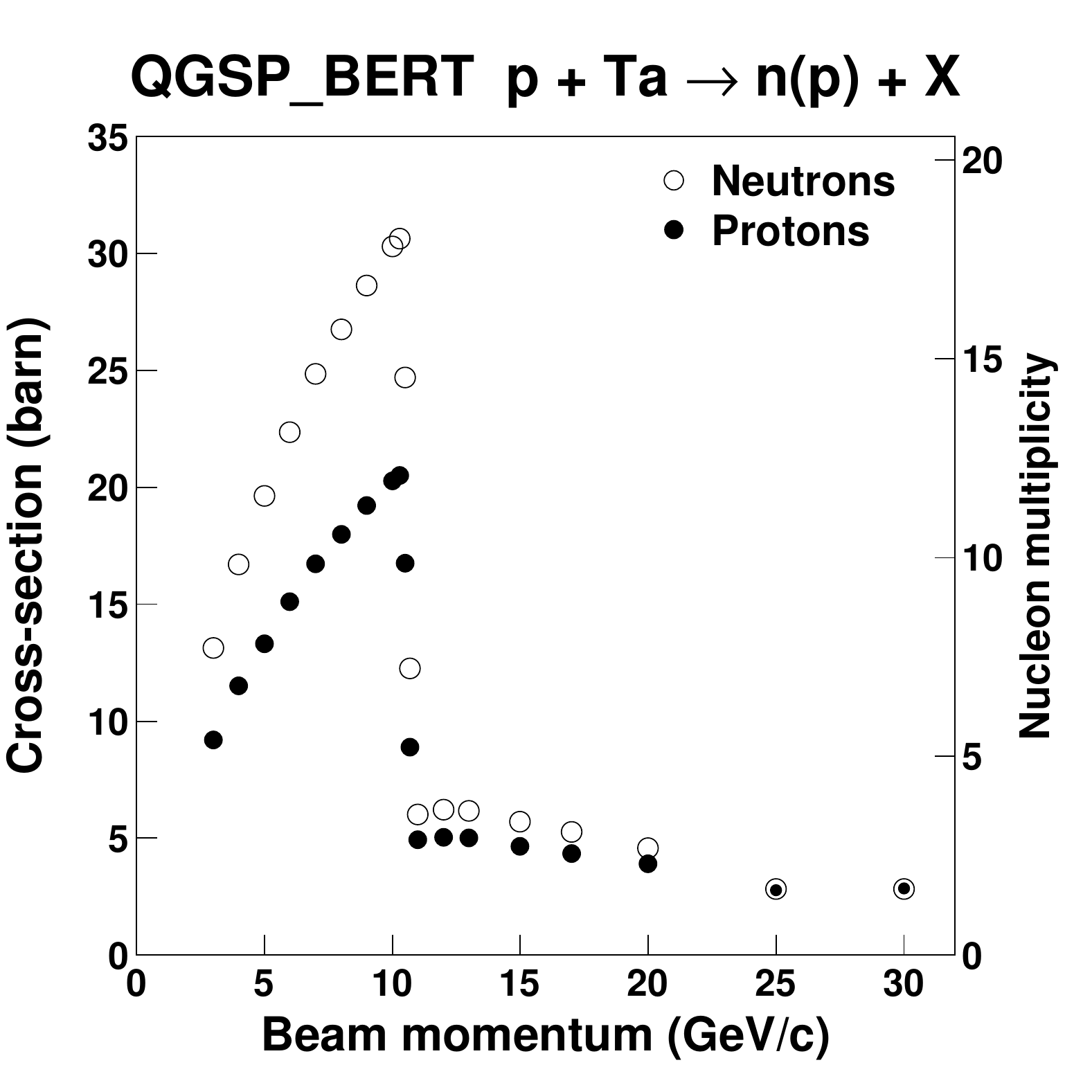} 
\caption{Geant4 simulation (QGSP\_BERT physics list) of the generation of 
secondary protons and neutrons with $p_{\rm T} > 0.1$~GeV/{\it c} in the
interaction of protons with Ta nuclei.}
\label{G4protonsandneutrons}
\end{center}
\end{figure}

With a view to exploring further this discontinuity around 10~GeV/{\it c} in Geant4, 
Fig.~\ref{G4protonsandneutrons} 
shows the 
simulation of the inclusive generation
of secondary neutrons and protons with a rather loose cut of $p_{\rm T} > 0.1$~GeV/{\it c},
in the interaction of beam protons with momentum in the range 3--30~GeV/{\it c} with Ta nuclei. Most prominent is an abrupt and unphysical change of modelling around 10~GeV/{\it c}. 
Further, a comparison with the large proton multiplicity shown in Fig.~\ref{G4protonsandneutrons}---up to some 20 protons per beam particle interaction---with the measured much smaller proton multiplicity (see, e.g., Ref.~\cite{Lead}) suggests that final-state protons are dominated by protons from the fragmentation of the target nucleus and not by final-state protons from the interactions of the 
primary beam particle or of secondary hadrons in the nuclear matter of the target nucleus. Since for protons from the fragmentation of the target nucleus it is expected that the momentum spectrum of secondary protons is steeply 
falling with increasing momentum, the cut $p_{\rm T} > 0.3$~GeV/{\it c} will strongly diminish the protons from the fragmentation of the target nucleus.

\begin{figure}[h]
\begin{center}
\begin{tabular}{c}
\includegraphics[width=0.45\textwidth]{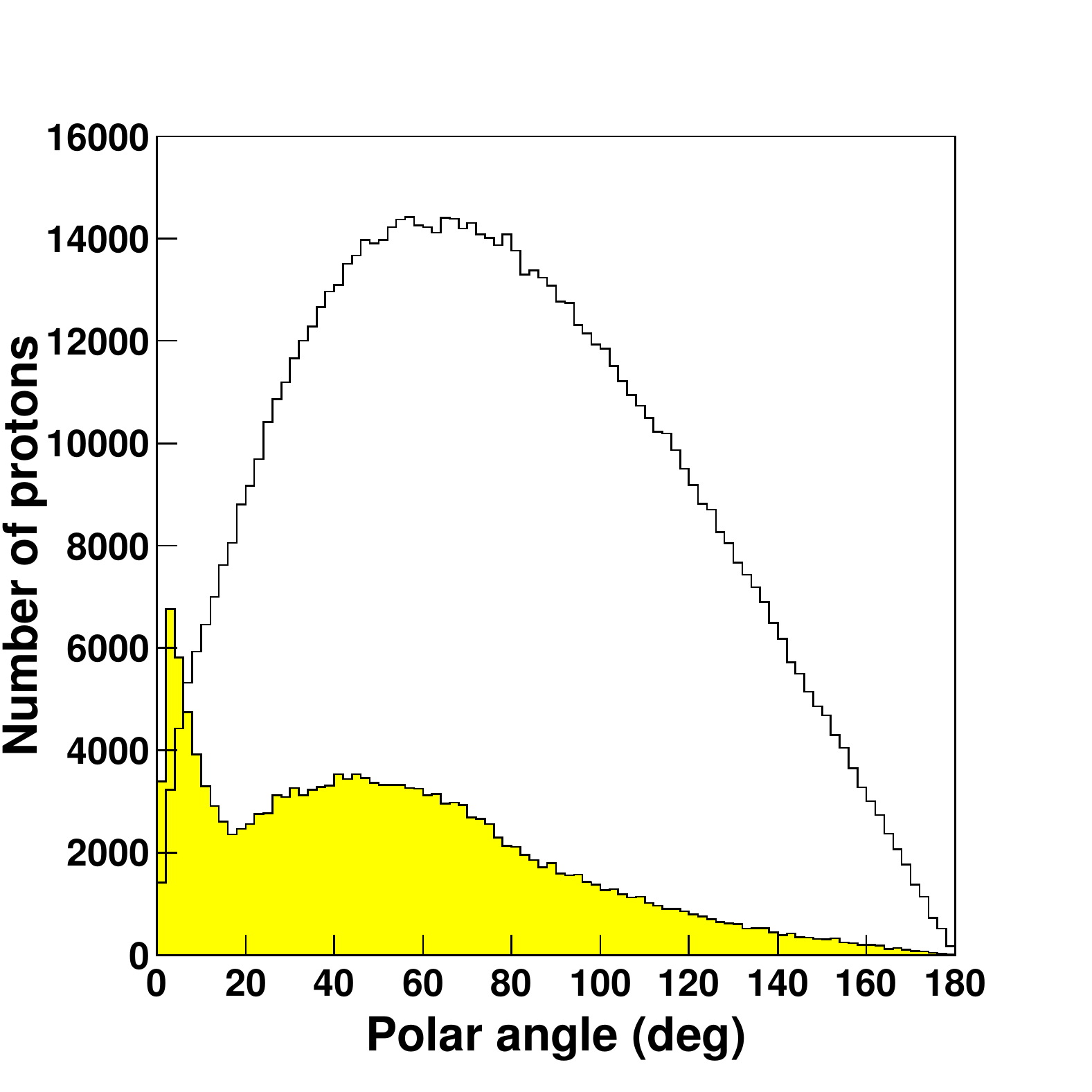} \\
\includegraphics[width=0.45\textwidth]{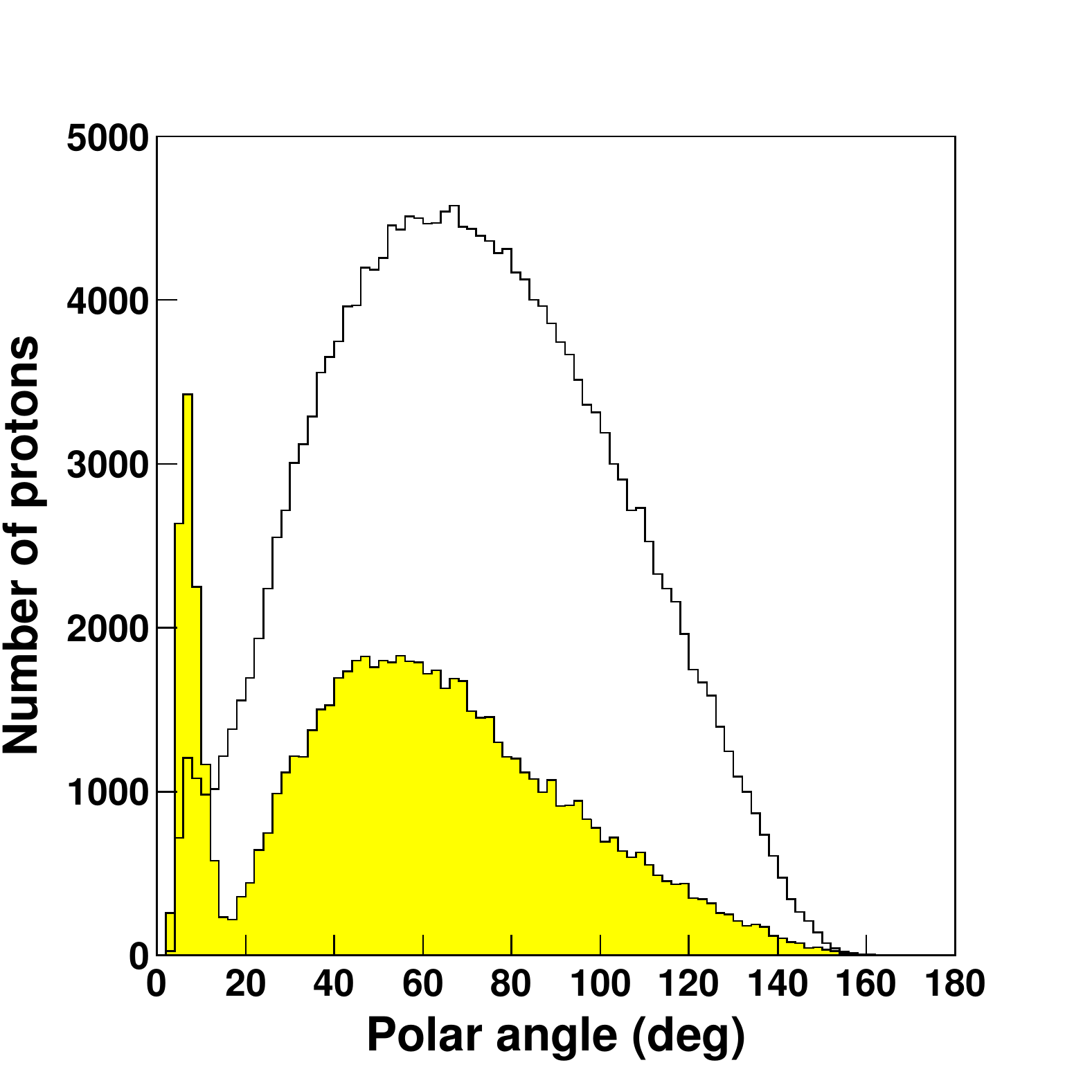} \\
\end{tabular}
\caption{The polar-angle distribution of protons generated by Geant 4 (QGSP\_BERT physics list) in the interaction of protons with Ta nuclei; the open histograms refer to proton beam momenta just below the discontinuity around 10~GeV/{\it c} beam momentum, the shaded (yellow)
histograms to proton beam momenta just above; the upper panel has no cut applied while the lower panel shows protons with $p_{\rm T} > 0.3$~GeV/{\it c}.} 
\label{FLUKAG4protonscutandnocut}
\end{center}
\end{figure}

Fig.~\ref{FLUKAG4protonscutandnocut} shows the polar-angle distribution of protons generated by Geant4 (QGSP\_BERT physics list), in the interaction of protons with Ta nuclei. The open histograms refer to proton beam momenta below the discontinuity around 10 GeV/{\it c}
beam momentum, the shaded histograms to proton beam momenta above. The upper panel has no cut applied while the lower panel shows protons with $p_{\rm T} > 0.3$~GeV/{\it c}. The discrepancy between the distributions below and above the discontinuity around 10~GeV/{\it c} is unphysical.

\begin{figure}[h]
\begin{center}
\begin{tabular}{c}
\includegraphics[width=0.45\textwidth]{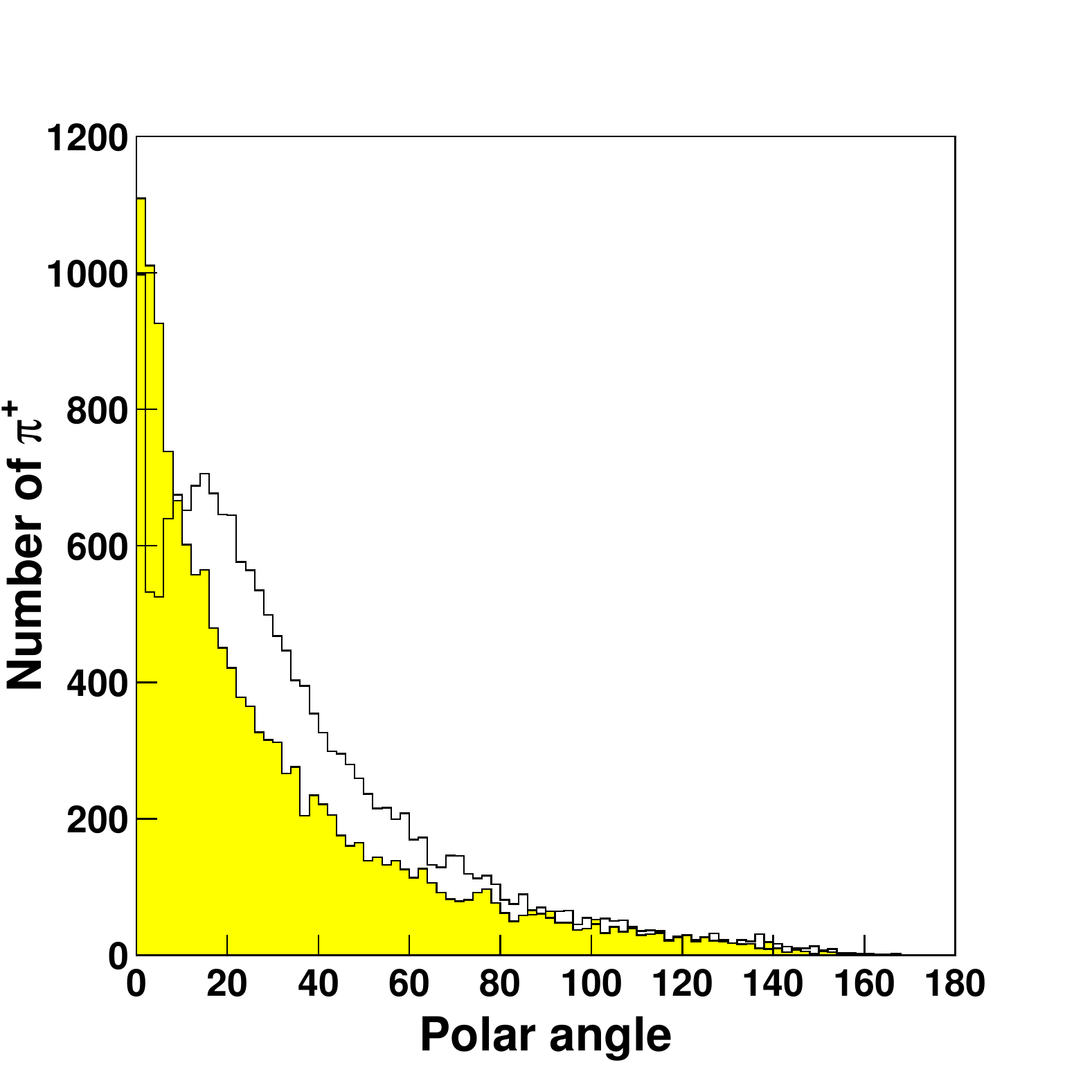} \\
\includegraphics[width=0.45\textwidth]{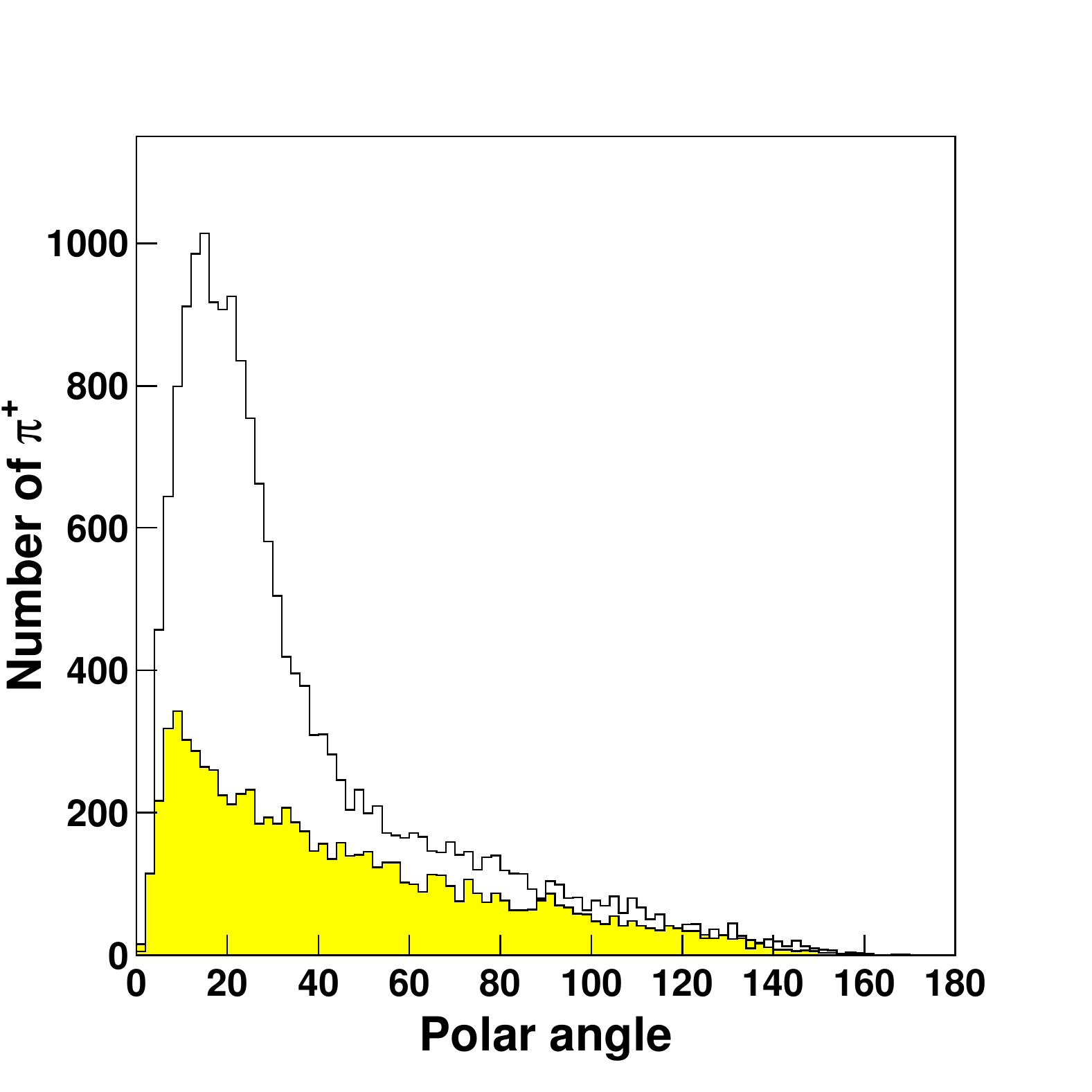} \\
\end{tabular}
\caption{The polar-angle distributions
of $\pi^+$'s generated by protons interacting with Ta nuclei, with 
no cuts applied; the upper panel shows the FLUKA simulation, the lower panel
the Geant4 (QGSP\_BERT physics list) simulation; the open histograms refer 
to proton beam momenta just above the discontinuities 
(around 5~GeV/{\it c} for FLUKA, around 10~GeV/{\it c} for Geant4),
the shaded (yellow) histograms to proton beam momenta just below.}
\label{FLUKAG4pionsinFlukaandGeant4}
\end{center}
\end{figure}

Figure~\ref{FLUKAG4pionsinFlukaandGeant4} shows the polar-angle distributions
of $\pi^+$'s generated by protons interacting with Ta nuclei. 
No cuts are applied. The upper panel shows the FLUKA simulation, the lower panel
the Geant4 (QGSP\_BERT physics list) simulation. Here the open histograms refer 
to proton beam momenta just above the discontinuities discussed before
(around 5~GeV/{\it c} for FLUKA, around 10~GeV/{\it c} for Geant4),
the shaded histograms to proton beam momenta just below. The differences
of the polar-angle distributions below and above the discontinuities
are unphysical.

Based on the same raw data from which we extracted our inclusive cross-sections, the HARP Collaboration published inclusive cross-sections of the production of $\pi^+$ and $\pi^-$ (but not of protons) by incoming proton, $\pi^+$ and $\pi^-$ beam particles, in the polar-angle
region $0.35 \; {\rm rad}  \leq \theta \leq 2.15 \; {\rm rad}$~\cite{OffLAprotonpaper,OffLApionpaper}, and compared their results with predictions of the Geant4 and MARS~\cite{MARS}  Monte Carlo tool kits. Substantial discrepancies were observed between their and our results, documented in Refs.~\cite{Beryllium1,Beryllium2,Copper, Tantalum,Lead}, discussed in Refs~\cite{JINSTpub,EPJCpub,WhiteBook1,WhiteBook2,WhiteBook3}, and summarized in the Appendix of Ref.~\cite{Beryllium1}.

\clearpage

\section{Synopsis}

A comprehensive comparison of measured inclusive cross-sections of proton, $\pi^+$
and $\pi^-$ production by beams of protons, $\pi^+$'s and $\pi^-$'s with
momentum in the range 3--15~GeV/{\it c}, interacting with beryllium, copper
and tantalum target nuclei, with simulations by the FLUKA and Geant4 Monte Carlo
tool kits is presented. Overall production cross-sections are reasonably well reproduced, within factors of two.
In more detail, there are areas with poor agreement that are unsatisfactory and call for modelling improvements. Overall, the current FLUKA simulation fares better than the current Geant4 simulation.

\section*{Acknowledgements}

We are greatly indebted to many technical collaborators whose 
diligent and hard work made the HARP detector a well-functioning 
instrument. We thank all HARP colleagues who devoted time and 
effort to the design and construction of the detector, to data taking, 
and to setting up the computing and software infrastructure. 
We express our sincere gratitude to HARP's funding agencies 
for their support.


\begin{thebibliography}{99}

\bibitem{Beryllium1} A.~Bolshakova {\it et al.}, Eur. Phys. J. {\bf C62} (2009) 293 
(CERN--PH--EP--2008--022, arXiv:0901.3648) 

\bibitem{Beryllium2} A.~Bolshakova {\it et al.}, Eur. Phys. J. {\bf C62} (2009) 697 
(CERN--PH--EP--2008--025, arXiv:0903.2145) 

\bibitem{Tantalum} A.~Bolshakova {\it et al.}, Eur. Phys. J. {\bf C63} (2009) 549
(CERN--PH--EP--2009--009, arXiv:0906.0471) 

\bibitem{Copper} A.~Bolshakova {\it et al.}, Eur. Phys. J. {\bf C64} (2009) 181
(CERN--PH--EP--2009--012, arXiv:0906.3653) 

\bibitem{Lead} A.~Bolshakova {\it et al.}, Eur. Phys. J. {\bf C66} (2010) 57
(CERN--PH--EP--2009--025, arXiv:0912.0378) 

\bibitem{GEANTpub} A.~Bolshakova {\it et al.},  Eur. Phys. J. {\bf C56} (2008) 323
(CERN--PH--EP--2008--007, arXiv:0804.3013) 

\bibitem{Geant4} S.~Agostinelli {\it et al.}, 
Nucl. Instrum. Methods Phys. Res. {\bf A506} (2003) 250; 
J.~Allison {\it et al.}, IEEE Trans. Nucl. Sci. {\bf 53} (2006) 270

\bibitem{ResponsetoGEANTpub} V.~Uzhinsky {\it et al.},
Eur. Phys. J. {\bf C61} (2009) 237

\bibitem{FLUKA} FLUKA: a multi-particle transport code,
A.~Ferrari {\it et al.},
CERN--2005--10 (INFN--TC--05--11, SLAC--R--773);
The FLUKA code: Description and benchmarking,
G.~Battistoni {\it et al.},
Proc. of the Hadronic Shower Simulation Workshop,
Fermilab (Btavia, Illinois), 6--8 September 2006 (eds. M.~Albrow and R.~Raja),
AIP Conf. Proc. {\bf 896} (2007) 31 

\bibitem{TPCpub} V.~Ammosov {\it et al.},
Nucl. Instrum. Methods Phys. Res. {\bf A588} (2008) 294
(CERN--PH--EP--2007--030)

\bibitem{RPCpub} V.~Ammosov {\it et al.},
Nucl. Instrum. Methods Phys. Res. {\bf A578} (2007) 119 
(CERN--PH--EP--2007--005)

\bibitem{Beryllium1tables} A.~Bolshakova {\it et al.}, CERN--HARP--CDP--2009--001  

\bibitem{Beryllium2tables} A.~Bolshakova {\it et al.}, CERN--HARP--CDP--2009--002

\bibitem{Tantalumtables} A.~Bolshakova {\it et al.}, CERN--HARP--CDP--2009--003

\bibitem{Coppertables} A.~Bolshakova {\it et al.}, CERN--HARP--CDP--2009--004

\bibitem{Leadtables} A.~Bolshakova {\it et al.}, CERN--HARP--CDP--2009--005

\bibitem{PreferredChoice} John Apostolakis {\it et al.},
Progress in hadronic physics modelling in Geant4,
J. Phys. Conf. Series {\bf 160} (2009) 012073

\bibitem{OffLAprotonpaper} M.G.~Catanesi {\it et al.}, Phys. Rev. {\bf C77} (2008) 055207
(arXiv:0805.2871)

\bibitem{OffLApionpaper} M.~Apollonio {\it et al.}, Phys. Rev. {\bf C80} (2009) 065207
(CERN--PH--EP--2009--021, arXiv:0907.1428)

\bibitem{MARS} N.V.~Mokhov, Report Fermilab-FN-628 (1995); 
N.V.~Mokhov and S.I.~Striganov, 
Proc. of the Hadronic Shower Simulation Workshop,
Fermilab (Btavia, Illinois), 6--8 September 2006 (eds. M.~Albrow and R.~Raja),
AIP Conf. Proc. {\bf 896} (2007) 50

\bibitem{JINSTpub} V.~Ammosov {\it et al.}, 
J. Instrum. {\bf 3} (2008) P01002

\bibitem{EPJCpub} V.~Ammosov {\it et al.},
Eur. Phys. J. {\bf C54} (2008) 169

\bibitem{WhiteBook1} V.~Ammosov {\it et al.}, 
CERN--HARP--CDP--2006--003

\bibitem{WhiteBook2} V.~Ammosov {\it et al.}, 
CERN--HARP--CDP--2006--007

\bibitem{WhiteBook3} V.~Ammosov {\it et al.}, 
CERN--HARP--CDP--2007--001

\end{thebibliography}
\end{document}